\documentclass[preprint,prc,a4paper,tightenlines,nofootinbib,unsortedaddress]{revtex4}
\usepackage{amssymb}
\usepackage{amsmath}
\usepackage{graphicx}
\usepackage{bm} 
\usepackage[loose]{subfigure}
\usepackage{xcolor}
\usepackage[squaren]{SIunits}
\usepackage{multirow}
\usepackage{braket}
\usepackage{tensor}
\usepackage{slashed}
\usepackage{array}

\newcommand{\di}{\mathrm{d}}

\newcommand{\punkt}{\;\text{.}}
\renewcommand{\vec}[1]{\boldsymbol{#1}}
\newcommand{\komma}{\;\text{,}}
\renewcommand{\i}{\mathrm{i}}

\begin{document}

\title{The reaction {\boldmath$\pi N\to\pi\pi N$} 
in chiral effective field theory\\[0.3em] 
with explicit {\boldmath$\Delta(1232)$ degrees of freedom}}

\author{D.~Siemens}
\affiliation{Institut f\"ur Theoretische Physik II, Ruhr-Universit\"at 
Bochum, D-44780 Bochum, Germany}

\author{V.~Bernard}
\affiliation{Institut de Physique Nucl\'eaire, CNRS/Univ. Paris-Sud
  11, (UMR 8608), F-91406 Orsay Cedex, France}

\author{E.~Epelbaum, H.~Krebs}
\affiliation{Institut f\"ur Theoretische Physik II, Ruhr-Universit\"at 
Bochum, D-44780 Bochum, Germany}


\author{Ulf-G.~Mei{\ss}ner}\
\affiliation{Helmholtz-Institut f\"ur Strahlen- und
             Kernphysik and Bethe Center for Theoretical Physics, \\
             Universit\"at Bonn,  D--53115 Bonn, Germany\\}%
\affiliation{Institute~for~Advanced~Simulation, Institut~f\"{u}r~Kernphysik,
J\"{u}lich~Center~for~Hadron~Physics, and JARA~-~High~Performance~Computing
Forschungszentrum~J\"{u}lich,
D-52425~J\"{u}lich, Germany}

%

\begin{abstract}
The reaction $\pi N\to\pi\pi N$ is studied at tree level up to
next-to-leading order in the 
framework of manifestly covariant  baryon chiral perturbation theory
with explicit $\Delta(1232)$ degrees of
freedom. Using total cross section data to determine the relevant low-energy
constants, predictions are made for various differential as well as
total cross sections at higher energies. A detailed comparison of
results based on the heavy-baryon and relativistic formulations of chiral perturbation theory 
with and without explicit $\Delta$ degrees of freedom is given. 
\end{abstract}

\maketitle

\section{Introduction}

Chiral perturbation theory ($\chi$PT) is nowadays a standard tool to analyze
low-energy hadronic reactions in harmony with the symmetries of
QCD. It was originally formulated by Weinberg \cite{Weinberg:1978kz} and, a few years
later, extended and applied by Gasser and Leutwyler to study the low-energy dynamics of
the Goldstone bosons at the one-loop level both in the SU(2) \cite{Gasser:1983yg}
and SU(3) \cite{Gasser:1984gg} sectors.  Starting with the pioneering work by
Gasser et al.~\cite{Gasser:1987rb}, $\chi$PT has also been extensively used in the baryon
sector, see Refs.~\cite{Bernard:1995dp,Bernard:2006gx,Bernard:2007zu}
for review articles and references therein. In the framework of $\chi$PT,
low-energy hadronic observables are calculated within the chiral
expansion, where  the expansion parameter $Q$ is defined as the ratio of
the soft scales corresponding to external momenta, denoted generically
by $q$, or the pion mass $M_\pi$, and the chiral symmetry breaking scale $\Lambda_\chi \sim 1$ GeV.  
While in the Goldstone boson
sector the hard scale $\Lambda_\chi$ only enters the amplitude through
values of the low-energy constants (LECs) so that pion loop integrals calculated using dimensional regularization
(DR) automatically fulfill the chiral power counting, 
a special treatment of the nucleon mass $m_N \sim \Lambda_\chi$ is required in
the baryon sector. The standard way to maintain the power counting in
the nucleon sector is the use of the heavy-baryon (HB)  version 
of the effective Lagrangian \cite{Jenkins:1990jv,Bernard:1992qa}. 
In the heavy-baryon formulation of chiral
perturbation theory (HB$\chi$PT), the nucleon mass does not appear in the
propagators and enters only in form of $1/m_N$-corrections to 
the vertices which leads to the same suppression of DR loop integrals
as in the Goldstone boson sector. It is, however, known that the HB
expansion has, for certain observables such as some of  the nucleon
electroweak and scalar form
factors  \cite{Bernard:1996cc,Becher:1999he}, a very limited range of
convergence. It is then advantageous to use a manifestly
Lorentz-invariant effective Lagrangian rather than its HB 
version. Power counting can still be maintained using the method of  
Becher and Leutwyler \cite{Becher:1999he} to extract the soft
(i.e. infrared singular) parts of the loop integrals leading to the
so-called infrared regularized 
$\chi$PT. Alternatively, the proper chiral scaling of the loop
integrals can be restored in the covariant framework by imposing 
the appropriate renormalization conditions as proposed in 
Refs.~\cite{Gegelia:1999gf,Fuchs:2003qc} within the so-called extended
on-mass-shell scheme (EOMS).  We refer the reader to Ref.~\cite{Bernard:2007zu} for
a detailed discussion and comparison of the various $\chi$PT formulations
and their applications in the single-baryon sector, see also
Ref.~\cite{Epelbaum:2012ua} for a recent application of these ideas in the
two-nucleon sector.  

Another popular idea to extend the range of applicability of $\chi$PT in the nucleon
sector is based on the explicit  treatment of the $\Delta$(1232), 
the close-by resonance with the excitation energy of $\Delta \equiv m_\Delta - m_N = 293 \mbox{
  MeV}$. All effects of the $\Delta$ in the standard approach are
encoded in LECs of  pion-nucleon interactions beyond 
leading order. The small excitation energy of the $\Delta$ and its strong
coupling to the  $\pi N$ system lead, however, to unnaturally large values of
certain LECs which can, potentially, spoil the convergence of the chiral expansion. 
One can, therefore, argue that the \emph{explicit}
inclusion of the $\Delta$ in $\chi$PT by treating the delta-nucleon
mass splitting as a soft scale will allow to resum a certain class of
important contributions and  improve the convergence as compared to 
the $\Delta$-less theory \cite{Jenkins:1991es,Hemmert:1997ye}. 
The improved convergence of HB$\chi$PT-$\Delta$ compared to the standard HB$\chi$PT
has indeed been confirmed for $\pi N$ scattering \cite{Fettes:2000bb},
proton Compton scattering, see \cite{Lensky:2012ag} and references therein,  
nuclear forces, see
e.g.~\cite{Kaiser:1998wa,Krebs:2007rh,Epelbaum:2007sq}, and other
processes, see Ref.~\cite{Bernard:2007zu} for a review.  It should,
however, be emphasized that the explicit inclusion of the $\Delta$
makes calculations in the covariant framework considerably more
involved and also leads to the appearance of additional LECs.  

In the present work we analyze in detail single pion production off nucleons
from threshold up to the delta resonance region using various
formulations of  
$\chi$PT. The reaction $\pi N \to \pi \pi
N$ has already attracted considerable interest on the experimental
and theoretical sides which, historically, goes back to the
possibility of using this process for the extraction of  the $\pi \pi$
scattering lengths (see Refs.~\cite{Beringer:1992ic,Bernard:1994wf,Bernard:1995gx} 
in the  context of $\chi$PT and a more general discussion in Ref.~\cite{Olsson:1995iy} with
references to earlier work). 
Single pion production off nucleons is also of special interest from the point of view of chiral
perturbation theory. First of all, it involves three pions in the
initial and final states so that one may expect for the scattering
amplitude to be strongly constrained by the chiral symmetry of QCD. 
It therefore provides an excellent testing ground for $\chi$PT. On the
other hand, the relatively high energies involved in this pion
production reaction and the proximity of resonances with a strong
coupling to the $\pi \pi N$ final state make the pursuit of a theoretical
description of this reaction rather challenging. One expects for this process
to be particularly well suited for studying the role of the $\Delta$
isobar, relativistic effects and unitarity and thus for testing various
available formulations of $\chi$PT. Indeed, in Ref.~\cite{Jensen:1997em}
a relativistic tree level calculation including the $\Delta$ and 
the Roper resonance with leading order pion-baryon vertices (i.e. 
dimension one couplings only) was performed and the appearing parameters 
were determined from other sources. The resulting total and differential cross
section data were generally well described, encouraging further studies
in baryon chiral perturbation theory.
Last but not least, it is worth mentioning 
that the reaction $\pi N \to \pi \pi N$ provides complementary
information to pion-nucleon scattering in the sense that it is 
sensitive to certain LECs  which cannot be
extracted from the $\pi N$ system. The most prominent example is
the LEC $d_{16}$ which governs the quark mass dependence of the
nucleon axial vector coupling constant. As a matter of fact, the lack
of knowledge of its precise value represents  one of the main sources
of theoretical uncertainty in chiral extrapolations of nuclear
observables
\cite{Epelbaum:2002gb,Epelbaum:2002gk,Beane:2002vs,Beane:2002xf,Berengut:2013nh,Epelbaum:2012iu,Epelbaum:2013wla}.

All these arguments provide a strong motivation to take a fresh look
at the 
reaction $\pi N \to \pi \pi N$ in the framework of $\chi$PT. In 
Refs.~\cite{Bernard:1995gx}  and  \cite{Fettes:1999wp}, it was
analyzed within HB$\chi$PT at tree- and leading one-loop
levels, respectively. A tree-level calculation based on the
relativistic pion-nucleon Lagrangian is reported in
Ref.~\cite{Bernard:1997tq}. The role of unitarity corrections in a
heavy baryon calculation was in particular stressed in Ref.~\cite{Mobed:2005av}.
While these studies already showed that
the predictions of chiral perturbation theory are in a satisfactory
agreement with experimental data, we expect to be able to improve on
them in the delta region by explicitly taking into account  the delta
degrees of freedom systematically, extending the earlier work of 
Ref.~\cite{Jensen:1997em}. To the best of our knowledge, no
calculations of this reaction using $\chi$PT with explicit $\Delta$'s
beyond the leading order pion-baryon couplings has been performed. 
In this paper we fill this gap and
study single pion production off nucleons in the framework of
relativistic baryon $\chi$PT with explicit $\Delta$
degrees of freedom at {\em complete} tree level 
with the inclusion of the terms from
the dimension-two effective Lagrangian.

Our paper is organized as
follows. In section \ref{sec:lagr} we discuss the relevant terms in
the effective pion-nucleon-delta Lagrangian. The decomposition of the
transition matrix elements into the corresponding invariant amplitudes
is considered in section \ref{sec:ampl} while the relevant observables
are defined in section \ref{sec:obs}. Section \ref{sec:graphs}
specifies all tree-level contributions to the amplitude up to
next-to-leading order.  The details of the calculation
and the fitting procedure can be found in section \ref{sec:fits}, while
predictions for observables not used in the fitting procedure are
collected in section \ref{sec:predictions}. Finally, the main results
of our study are summarized in section \ref{sec:sum}. The appendix
contains explicit expressions for the kinematical variables and weight
functions we are using.

\section{Effective Lagrangian}
\label{sec:lagr}

We employ the so-called small scale expansion (SSE) or $\varepsilon$-expansion 
throughout this work with the expansion parameter being defined as  \cite{Hemmert:1997ye}
\begin{equation}
  \label{eq:1}
  \varepsilon\in \left\{\frac{q}{\Lambda_\chi},\frac{M_\pi}{\Lambda_\chi},\frac{\Delta}{\Lambda_\chi} \right\}\,,
\end{equation} 
i.e.~the delta-nucleon mass splitting is treated on the same footing
as the pion mass, see however Ref.~\cite{Pascalutsa:2002pi} for an alternative power
counting scheme.   We now discuss the terms in the effective
Lagrangian relevant for our calculation.

The relativistic effective Lagrangian needed to describe pion-nucleon
dynamics at tree level consists 
of the following pieces (see Ref.~\cite{Fettes:2000gb} for a full list of terms)
\begin{equation}
  \label{eq:2}
  \mathcal{L}_\mathrm{eff}=\mathcal{L}_{\pi\pi}^{(2)}
    +\mathcal{L}_{\pi N}^{(1)}+\mathcal{L}_{\pi N}^{(2)}+
    \mathcal{L}_{\pi \Delta}^{(1)}+\mathcal{L}_{\pi \Delta}^{(2)}+
    \mathcal{L}_{\pi N\Delta}^{(1)}+\mathcal{L}_{\pi N\Delta}^{(2)}\komma
\end{equation}
where the superscripts refer to the chiral dimension. 
The first term in Eq. \eqref{eq:2} describes the meson interaction
\begin{align}
  \label{eq:3}
      \mathcal{L}_{\pi\pi}^{(2)}&=\frac{F_\pi^2}{4}\braket{\partial_\mu
    U^\dagger \partial^\mu U}+\frac{F_\pi^2B}{2}\braket{\mathcal{M}
    U^\dagger+ U \mathcal{M}}\komma
\end{align}
where the pions are collected in the SU(2) matrix-valued field
$U(x)=u(x)^2$  given by 
\begin{equation}
  \label{eq:4}
U = 1 + \i \frac{\vec{\tau} \cdot \vec{\pi}}{F_\pi} - \frac{\vec{\pi}^2}{2 F_\pi^2} - \i \alpha  \frac{
\vec{  \pi}^2 \vec{\tau} \cdot \vec{\pi} }{F_\pi^3} 
+ \frac{(8 \alpha - 1)}{8 F_\pi^4} \vec{\pi}^4 + \ldots \,,
\end{equation}
where $\alpha$ is a constant representing the freedom in the
definition of the pion field. 
Further, $F_\pi$ is the pion decay constant\footnote{Since the
  calculation in the present work is carried out at the tree level, we
do not have to differentiate between the pion decay constant in the
chiral limit and its physical value. The same applies also to other
quantities such as the nucleon mass and the nucleon axial vector
coupling. }, $\mathcal{M} = {\rm diag} (m_u, \, m_d)$ is
the quark mass matrix and $B$ is a low-energy constant.   

The next two terms in Eq.~(\ref{eq:2}) give the leading and subleading
pion-nucleon Lagrangians
\begin{align}
  \label{eq:6}
  \begin{aligned}
    \mathcal{L}_{\pi N}^{(1)}&=\bar{\Psi} \left[ \i
      \slashed{D}-m_N+\frac{g_A}{2}\slashed{u}\gamma_5 \right]\Psi\komma\\
    \mathcal{L}_{\pi N}^{(2)}&= \bar{\Psi} \left[c_1\Braket{\chi_+} +
      c_2\left(-\frac{1}{8m_N^2}\braket{u_\mu
          u_\nu}\{D^\mu,D^\nu\}+\mathrm{h.c.} \right)+
      \frac{c_3}{2}\Braket{u\cdot u}\right.\\
    &\qquad\left.-\frac{c_4}{8} [\gamma^\mu, \gamma^\nu][u_\mu, u_\nu]
      + c_5 \Big( \chi_+ -\frac{\braket{\chi_+}}{2} \Big)\right]
    \Psi\komma
  \end{aligned}
\end{align}
where $m_N$ and $g_A$ denote the nucleon mass and axial vector
coupling, $c_i$'s refer to further LECs and  the proton and neutron are given in the isodoublet representation
\begin{equation}
  \label{eq:5}
  \Psi=
  \begin{pmatrix}
    p\\n 
  \end{pmatrix}
\punkt
\end{equation}
The covariant derivative in Eq. \eqref{eq:6}  in the
absence of external sources is defined via:
\begin{equation}
  \label{eq:9}
  D_\mu = \partial_\mu + \Gamma_\mu\
\qquad\mathrm{with}\qquad
  \Gamma_\mu=\frac{1}{2}(u^\dagger\partial_\mu u +u \,\partial_\mu u^\dagger)\punkt
\end{equation}
In addition, the abbreviation
\begin{align}
  \label{eq:10}
  \chi_+ = u^\dagger \chi u^\dagger + u\chi^\dagger u \qquad
  \mathrm{with} \qquad \chi= 2B_0 \mathcal{M}
\end{align}
and the chiral vielbein
\begin{equation}
  \label{eq:7}
  u_\mu= \i ( u^\dagger\partial_\mu u-u\,\partial_\mu u^\dagger )
\end{equation}
are used. 

The inclusion of the $\Delta(1232)$ as an explicit degree of freedom
adds the four last terms to the effective Lagrangian  in
Eq.~(\ref{eq:2}). The delta isobar is
described by a
Rarita-Schwinger isospurion $\Psi_\mu^i\komma$ a spin-$3/2$
field, which is constructed via coupling of a spin-$1$ to a
spin-$1/2$ field,
treated as an isodoublet with an additional isovector index $i\in\{1,
2, 3\}$. The pion-delta
Lagrangian up to second order reads \cite{Hemmert:1997ye,Fettes:2000bb} 
\begin{align}
  \label{eq:13}
  \begin{aligned}
    \mathcal{L}_{\pi\Delta}^{(1)} &=- \bar{\Psi}_i^\mu\Big[
    (\i\slashed{D}^{ij}-m_\Delta \delta^{ij} )g_{\mu\nu}-\i (\gamma_\mu
    D_\nu^{ij}+\gamma_\nu D_\nu^{ij})+\i \gamma_\mu
    \slashed{D}^{ij}\gamma_\nu+m_\Delta \delta^{ij}\gamma_\mu \gamma_\nu \\
    &\qquad+\frac{g_1}{2}g_{\mu\nu} \slashed{u}^{ij} \gamma_5 +
    \frac{g_2}{2}(\gamma_\mu
    u_\nu^{ij}+u_\mu^{ij}\gamma_\nu)\gamma_5+\frac{g_3}{2}\gamma_\mu
    \slashed{u}^{ij}\gamma_5 \gamma_\nu
    \Big]\Psi^\nu_j\komma\\
    \mathcal{L}_{\pi\Delta}^{(2)}&=\frac{c_1^\Delta}{2}
    \bar{\Psi}_\mu^i\Theta^{\mu\alpha}
    (z)\gamma_{\alpha\beta}\delta^{ij}\braket{\chi_+}\Theta^{\beta\nu}(z)\Psi_\nu^j
    +\mathrm{h.c.} + \ldots \,,
  \end{aligned}
\end{align}
where terms with two and more pion fields in
$\mathcal{L}_{\pi\Delta}^{(2)}$ are not shown since they do not contribute to the
reaction $\pi N \to \pi \pi N$ at order $\varepsilon^2$.
Here,  the quantity  $\Theta^{\mu\alpha}(z)$ is defined via $\Theta^{\mu\alpha}(z)=g^{\mu\alpha}+z\gamma^\mu\gamma^\alpha$
and $z$, $g_2$ and $g_3$ denote the so-called off-shell
parameters. 
Notice that the dependence of the amplitude on such
off-shell parameters can be  absorbed into a redefinition of the
corresponding LECs, see
Refs.~\cite{Tang:1996sq,Pascalutsa:2000kd,Krebs:2009bf} for more
details. It is, therefore, convenient
to choose
\begin{equation}
  \label{eq:200}
  \begin{aligned}
    z=g_2=g_3=0 , 
  \end{aligned}
\end{equation}
which specifies our conventions for the calculations in the manifestly
covariant framework.   
Here and in what follows, we show explicitly the dependence on some of the
off-shell parameters in order to maintain consistency between the
covariant and HB calculations as will be explained below.  
The covariant
derivative in Eq. \eqref{eq:13} is given by
\begin{equation}
  \label{eq:14}
  D^{ij}_\mu=\partial_\mu \delta^{ij}+\Gamma_\mu^{ij}
\qquad\mathrm{with}\qquad
  \Gamma_\mu^{ij}=\Gamma_\mu\delta^{ij} - \i \epsilon^{ijk} \braket{\tau^k\Gamma_\mu}\punkt
\end{equation}
Finally, the pion-nucleon-delta Lagrangian reads  \cite{Hemmert:1997ye,Fettes:2000bb} 
\begin{align}
  \label{eq:16}
    \mathcal{L}_{\pi N \Delta}^{(1)}&= h_{A}\left[\bar{\Psi}_\mu^i\Theta^{\mu\alpha}(z)w_\alpha^i\Psi_N+\bar{\Psi}w^i_\alpha\Theta^{\alpha\mu}(z)\Psi_\mu^i\right]\komma
    \\
    \mathcal{L}_{\pi N \Delta}^{(2)}&= \bar{\Psi}_\mu^i\Theta^{\mu
      \alpha}(z)\left[ \i b_3 w_{\alpha\beta}^i\gamma^\beta +
      \frac{b_4}{2}w_\alpha^i w_\beta^j \gamma^\beta\gamma_5\tau^j +
      \frac{b_5}{2}w_\alpha^j w_\beta^i \gamma^\beta\gamma_5\tau^j +
      \frac{b_6}{m_N} \i w_{\alpha\beta}^i \i D^\beta \right]\Psi
    +\mathrm{h.c.},\nonumber
\end{align}
where $h_{A}$ is the $\pi N \Delta$ axial coupling, $b_i$ are further LECs and   
\begin{equation}
  \label{eq:17}
  \begin{aligned}
    w_\alpha^i &= \frac{1}{2}\mathrm{Tr}\left[ \tau^i u_\alpha
    \right] \qquad \mathrm{and} \qquad
    w_{\alpha\beta}^i &=\frac{1}{2}\mathrm{Tr}\left[ \tau^i\left[D_\alpha,u_\beta\right] \right]\punkt
  \end{aligned}
\end{equation}
It should be emphasized that  the free spin-3/2 Lagrangian is
non-unique and usually written in terms of an unphysical ``gauge'' parameter
$A$, whose entire dependence of the observables can be absorbed into 
redefinition of the delta field.  We have not shown explicitly the dependence on
the parameter $A$ in the effective Lagrangian by making the choice
$A=-1$. This particular choice is convenient in the covariant approach
since it leads to the simplest form of the free Lagrangian and thus
also of
the free delta propagator 
\begin{equation}
  \label{eq:18}
\mathcal{G}^{\mu\nu}_\Delta (p)= -\frac{\slashed{p}+m_\Delta}{p^2-m_\Delta^2}\left(
      g^{\mu\nu}-\frac{1}{3}\gamma^\mu \gamma^\nu
      +\frac{1}{3}\frac{p^\mu \gamma^\nu-p^\nu \gamma^\mu}{m_\Delta}
      -\frac{2}{3}\frac{p^\mu p^\nu}{m_\Delta^2}\right)
\punkt
\end{equation}

Since we are particularly interested here in the role of relativistic effects, we will also carry out the calculations 
within HB$\chi$PT.  In this approach, 
the nucleon four-momentum $p_\mu$ is separated into a large piece
close to the
on-shell kinematics and a soft residual contribution $k_\mu$ via
\begin{equation}
  \label{eq:19}
  p_\mu=m_N\, v_\mu +k_\mu
\end{equation}
with $v_\mu$ being the four-velocity of the nucleon with  the
properties
$v^2=1$ and $ v^0\geq 1$.
The nucleon field $\Psi$ is decomposed into eigenstates of
$\slashed{v}$ with eigenvalues $+1$ and $-1$, the so-called ''light''
and ''heavy'' fields, respectively,
\begin{equation}
  \label{eq:20}
  \begin{aligned}
    N_v &= e^{\i m_N v\cdot x}P_v^+\Psi\komma\\
    h_v &=  e^{\i m_N v\cdot x}P_v^-\Psi
  \end{aligned}
\end{equation}
with the projection operators $P_v^\pm=\frac{1}{2}(1\pm\slashed{v})$.
$N_v$ and $h_v$ correspond to the upper- and lower-components
of a Dirac spinor and thus positive and negative energy solutions,
respectively. The effects of the $h_v$-components  at low energies
can be interpreted as contact terms so that the
resulting Lagrangian involves only $N_v$ and its
derivatives. Analogously, the delta resonance is included by defining
a "light" spin-$3/2$ and isospin-$3/2$ field
\begin{align}
  \label{eq:21}
  T^\mu_i &= e^{\i m_N
    v\cdot x} P^+_v
  \xi^{3/2}_{ij}\left(P^{3/2}_{33}\right)^{\mu\nu}\Psi_\nu^j 
\end{align}
with the spin and isospin projection operators $P^{3/2}_{33}$ and
$\xi^{3/2}_{ij}$, respectively. The other ``heavy'' component is 
integrated out of the action. For more details on the
heavy-baryon expansion in the pion-nucleon-delta sector the reader is
referred to Refs.~\cite{Hemmert:1997ye,Fettes:2000bb}.  For the
calculation at order $\epsilon^2$,  the required heavy-baryon
effective Lagrangian involves the following pieces
\begin{equation}
  \label{eq:22}
  \begin{aligned}
    \mathcal{\hat{L}}_\mathrm{eff}&=\mathcal{L}_{\pi\pi}^{(2)}
    +\mathcal{\hat{L}}_{\pi N}^{(1)}+\mathcal{\hat{L}}_{\pi N}^{(2)}+
    \mathcal{\hat{L}}_{\pi \Delta}^{(1)}+\mathcal{\hat{L}}_{\pi \Delta}^{(2)}+
    \mathcal{\hat{L}}_{\pi N\Delta}^{(1)}+\mathcal{\hat{L}}_{\pi
      N\Delta}^{(2)}.  
  \end{aligned}
\end{equation}
The explicit form of the nucleon terms can be found  in
\cite{Fettes:2000gb} while  
the delta terms are given in Ref.~\cite{Hemmert:1997ye}. We emphasize,
however, that
the authors of Ref.~\cite{Hemmert:1997ye}, whose results for the HB
effective Lagrangian are adopted in our work, made a choice for the gauge
parameter $A=0$ without specifying the values of the off-shell
parameters. In order to be consistent with the convention used in the
covariant calculations, see Eq.~(\ref{eq:200}), one has to choose in the HB framework
\begin{equation}
  \label{eq:201}
  \begin{aligned}
    \hat{z}=-\frac{1}{2}\komma\qquad \hat{g}_2=-g_1\komma\qquad
    \hat{g}_3=-g_1 \,,
  \end{aligned}
\end{equation}
see Ref.~\cite{Krebs:2013} for more details.

\section{Invariant Amplitudes}
\label{sec:ampl}

In this section we discuss the decomposition of the $T$-matrix for the
reaction $\pi N \to \pi \pi N$ in terms of the corresponding invariant
amplitudes, following Ref.~\cite{Bernard:1997tq}. 
Throughout this work, the kinematical variables are defined as follows:
\begin{equation}
  \label{eq:24}
  \pi^a(q_1) \, N(p=m_N v +k) \; \to \; \pi^b(q_2)\, \pi^c(q_3) \, N^\prime(p^\prime=m_N v +k^\prime)\komma
\end{equation}
where $N$ denotes a nucleon and $\pi^a$ a
pion with the isospin quantum number $a$.

\subsection{Relativistic chiral perturbation theory}
In the relativistic case, the $T$-matrix can be expressed in
 terms of four invariant amplitudes $F_i$
($i\in\{1,2,3,4\}$) which depend on the five Mandelstam variables
\begin{equation}
  \label{eq:25}
    s=(p+q_1)^2\komma\quad s_1=(q_2+p^\prime)^2\komma\quad
  s_2=(q_3+p^\prime)^2\komma\quad t_1=(q_2-q_1)^2\komma\quad t_2=(q_3-q_1)^2\punkt
\end{equation}
The spin structure of the $T$-matrix can be parametrized in the
following way
\begin{equation}
  \label{eq:26}
  T_{ss^\prime}^{abc}=\i \bar{u}^{(s^\prime)}\gamma_5\left(F_1^{abc}+(\slashed{q}_2+\slashed{q}_3)F_2^{abc}
  +(\slashed{q}_2-\slashed{q}_3)F_3^{abc}
 +(\slashed{q}_2\slashed{q}_3-\slashed{q}_3\slashed{q}_2)F_4^{abc}\right) u^{(s)}, 
\end{equation}
where the superscripts on the spinors $\bar u, u$ refer to the spin.
The isospin decomposition of the invariant amplitudes reads
\begin{equation}
  \label{eq:27}
  F_i^{abc}=
\chi_{N^\prime}^\dagger\left(\tau^a\delta^{bc}B_i^1
+\tau^b\delta^{ac}B_i^2+\tau^c\delta^{ab}B_i^3
+\i\epsilon^{abc} B_i^4 \right)\chi_{N} . 
\end{equation}
Here,  the $B_i$'s have the following symmetry under exchange of
the two outgoing particles
\begin{equation}
  \label{eq:28}
  B_i^2 (s,s_1,s_2,t_1,t_2) = \epsilon_i B_i^3
  (s,s_2,s_1,t_2,t_1)\komma \qquad
  \epsilon_{1,2} =1\komma \qquad
  \epsilon_{3,4} =-1\punkt
\end{equation}
In the five physically accessible channels, the
amplitudes contributing to each channel reduce to
\begin{align}
\label{eq:29}
  \begin{aligned}
    &\mathrm{I.} &\pi^-p&\to \pi^0\pi^0n:   &F_i&=\sqrt{2}B_i^1\\
    &\mathrm{II.} &\pi^-p&\to \pi^+\pi^-n: &F_i&=\sqrt{2}(B_i^1+B_i^2)\\
    &\mathrm{III.} &\pi^+p&\to \pi^+\pi^+n: &F_i&=\sqrt{2}(B_i^2+B_i^3)\\
    &\mathrm{IV.} &\pi^+p&\to \pi^+\pi^0p: &F_i&=B_i^3+B_i^4\\
    &\mathrm{V.} &\pi^-p&\to \pi^0\pi^-p: &F_i&=B_i^2+B_i^4\punkt
  \end{aligned}
\end{align}
The unpolarized invariant matrix element squared in the relativistic
formalism has the form 
\begin{equation}
  \label{eq:31}
    |\mathcal{M}|^2=\frac{1}{2}\sum_{s,s^\prime}T^\dagger_{s
      s^\prime}T^{}_{s
      s^\prime}=\sum_{i,j=1}^4\frac{y_{ij}}{4m_N^2}F_i^*F_j , 
\end{equation}
with the weight functions $y_{ij}=y_{ji}$ given by the trace over the
respective Dirac structures (see Appendix \ref{sec:KinWeiFunc}).

\subsection{Heavy-baryon chiral perturbation theory}
In the heavy-baryon framework, the spin
structure of an amplitude is given by a combination of the
non-commuting Pauli-Lubanski spin vectors $S_\mu$. In the case of
$\pi N\to\pi\pi N$, the transition matrix can be written in terms of four
invariant amplitudes $A$, $B$, $C$ and $D$ which depend on the five
momenta $k$, $k^\prime$, $q_1$, $q_2$ and $q_3$ and are defined via \cite{Fettes:1999wp}
\begin{equation}
  \label{eq:32}
  \begin{aligned}
    T_{s s^\prime}^{abc} =\bar{u}_v^{(s^\prime)}
\left(
S\cdot q_1\,A^{abc}+S\cdot q_2\,B^{abc}+ S\cdot q_3\,C^{abc} + \i
\epsilon_{\mu\nu\alpha\beta}\,q_1^\mu q_2^\nu q_3^\alpha v^\beta\,D^{abc}
\right)u_v^{(s)}.
  \end{aligned}
\end{equation}
Here, the heavy-baryon spinor $u_v^{(s)}$ is given in the Pauli spinor
representation
\begin{equation}
  \label{eq:33}
  u_v^{(s)}(p)=P_v^+ u^{(s)}(p)=\mathcal{N}
\begin{pmatrix}
  \chi_s\\  0
\end{pmatrix}
\punkt
\end{equation}
The normalization factor
\begin{equation}
  \label{eq:34}
  \mathcal{N}=\sqrt{\frac{p^0+m_N}{2m_N}}
\end{equation}
ensures the proper matching to the relativistic theory and has
to be taken into account in the $1/m_N$ expansion,  
\begin{equation}
  \label{eq:35}
\mathcal{N}\mathcal{N}^\prime =1 +\mathcal{O}\left(\frac{1}{m_N^2}\right)\punkt
\end{equation}
Thus, for a tree level calculation, the normalization factors can be set equal
to $1$.
The isospin decomposition is the same as in the relativistic case, namely
\begin{equation}
  \label{eq:36}
  X^{abc}=
\chi_{N^\prime}^\dagger\left(\tau^a\delta^{bc}X_1+\tau^b\delta^{ac}X_2+\tau^c\delta^{ab}X_3+\i\epsilon^{abc}
    X_4 \right)\chi_{N}\komma\quad X\in\{A,B,C,D \}
\end{equation}
and thus the reduction in the five physically accessible channels is
the same
\begin{equation}
  \label{eq:37}
  \begin{aligned}
    &\mathrm{I.} &\pi^-p&\to \pi^0\pi^0n:   &X&=\sqrt{2}X_1 \\
    &\mathrm{II.} &\pi^-p&\to \pi^+\pi^-n: &X&=\sqrt{2}(X_1+X_2)\\
    &\mathrm{III.} &\pi^+p&\to \pi^+\pi^+n: &X&=\sqrt{2}(X_2+X_3)\\
    &\mathrm{IV.} &\pi^+p&\to \pi^+\pi^0p: &X&=X_3+X_4\\
    &\mathrm{V.} &\pi^-p&\to \pi^0\pi^-p: &X&=X_2+X_4\punkt
  \end{aligned}
\end{equation}
The unpolarized invariant matrix element squared in the heavy-baryon
formalism reads
\begin{equation}
  \label{eq:38}
  \begin{aligned}
    |\mathcal{M}|^2&=\frac{1}{2}\sum_{s,s^\prime}T^\dagger_{s
      s^\prime}T^{}_{s s^\prime}\\
&=\frac{1}{4}
\Big[
|A|^2\vec{q}_1^2+|B|^2 \vec{q}_2^2+ |C|^2 \vec{q}_3^2+ (A^* B+ A
B^*)\vec{q}_1\cdot \vec{q}_2
+ (A^* C+ A C^*)\vec{q}_1\cdot\vec{q}_3\\
&+
(B^* C+ B C^*)\vec{q}_2\cdot\vec{q}_3+ 4|D|^2 \vec{q}_1^2\vec{q}_2^2\vec{q}_3^2(1-x_1^2)(1-x_2^2)
\Big(1-\frac{(z-x_1x_2)^2}{(1-x_1^2)(1-x_2^2)}\Big)\Big]\komma
  \end{aligned}
\end{equation}
where $x_1$, $x_2$ and $z$ are the cosines of the angles between
$\vec q_1$ and $\vec q_2$, $\vec q_1$ and $\vec q_3$, and $\vec q_2$
and $\vec q_3$, respectively.

\section{Tree-level contributions to the scattering amplitude}
\label{sec:graphs}

\begin{figure}[ht]
  \centering
\includegraphics[width=\textwidth]{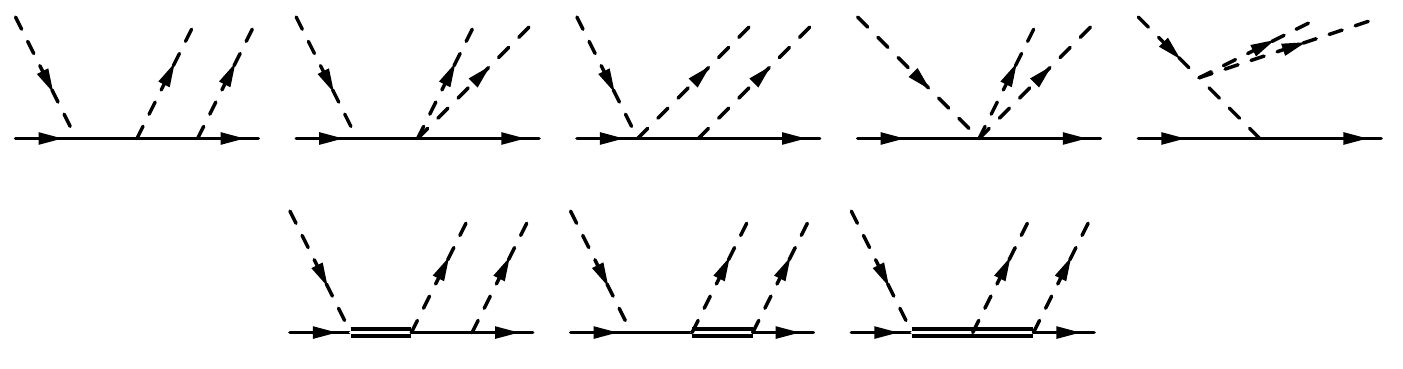}
  \caption{LO graphs for the reaction $\pi N\to\pi\pi N$.
 Nucleons and pions are denoted by solid and dashed
    lines, respectively. Delta is denoted by a double solid line. Crossed
    diagrams are not shown.}
  \label{fig:LOgraphsRel}
\end{figure}

\begin{figure}[ht]
  \centering
\includegraphics[width=\textwidth]{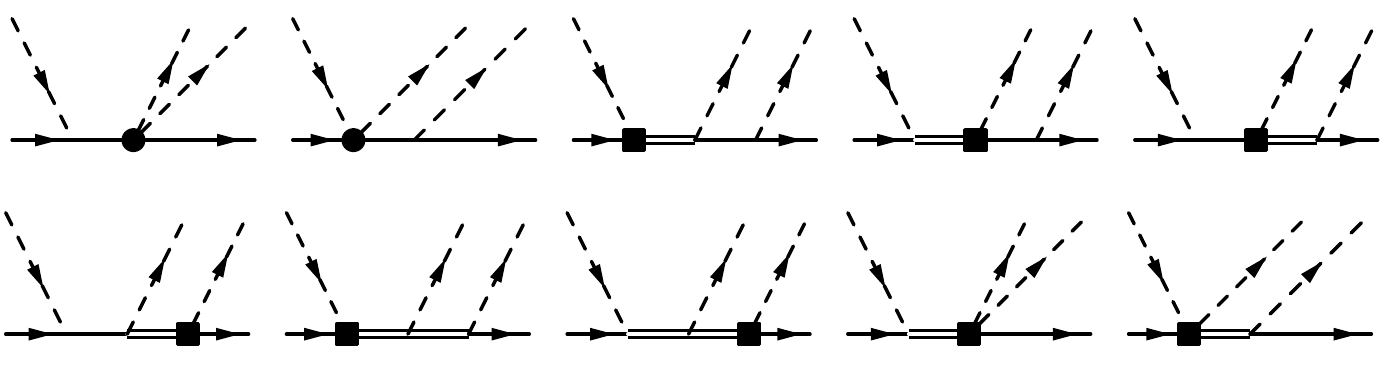}
  \caption{NLO graphs for the reaction $\pi N\to\pi\pi N$. The filled
    blob (filled square) denotes an insertion of the $c_i$- ($b_i$-) vertices. Crossed
    diagrams are not shown.}
  \label{fig:NLOgraphsRel} 
\end{figure}

The leading-order (LO)  and next-to-leading
order (NLO) diagrams emerging at orders $\varepsilon^1$
and $\varepsilon^2$ in the SSE  in the relativistic framework are shown in Fig.~\ref{fig:LOgraphsRel} 
and  Fig.~\ref{fig:NLOgraphsRel}, respectively.  
The LO diagrams are constructed solely from the lowest-order vertices
and thus depend only on the well-known LECs $F_\pi$, $g_A$
and the pion-nucleon-delta axial constant $h_{A}$. Subleading
diagrams involve a single insertion of the LECs $c_i$ from $\mathcal{L}_{\pi N}^{(2)}$, which are known from
pion-nucleon scattering or $b_i$ from  $\mathcal{L}_{\pi N
  \Delta}^{(2)}$. We do not show diagrams involving an insertion of
the LEC $c_1^{\Delta}$ whose contributions are taken into account by
using the physical values of the mass of the  delta isobar. 
 
\begin{figure}[ht]
  \centering
\includegraphics[width=\textwidth]{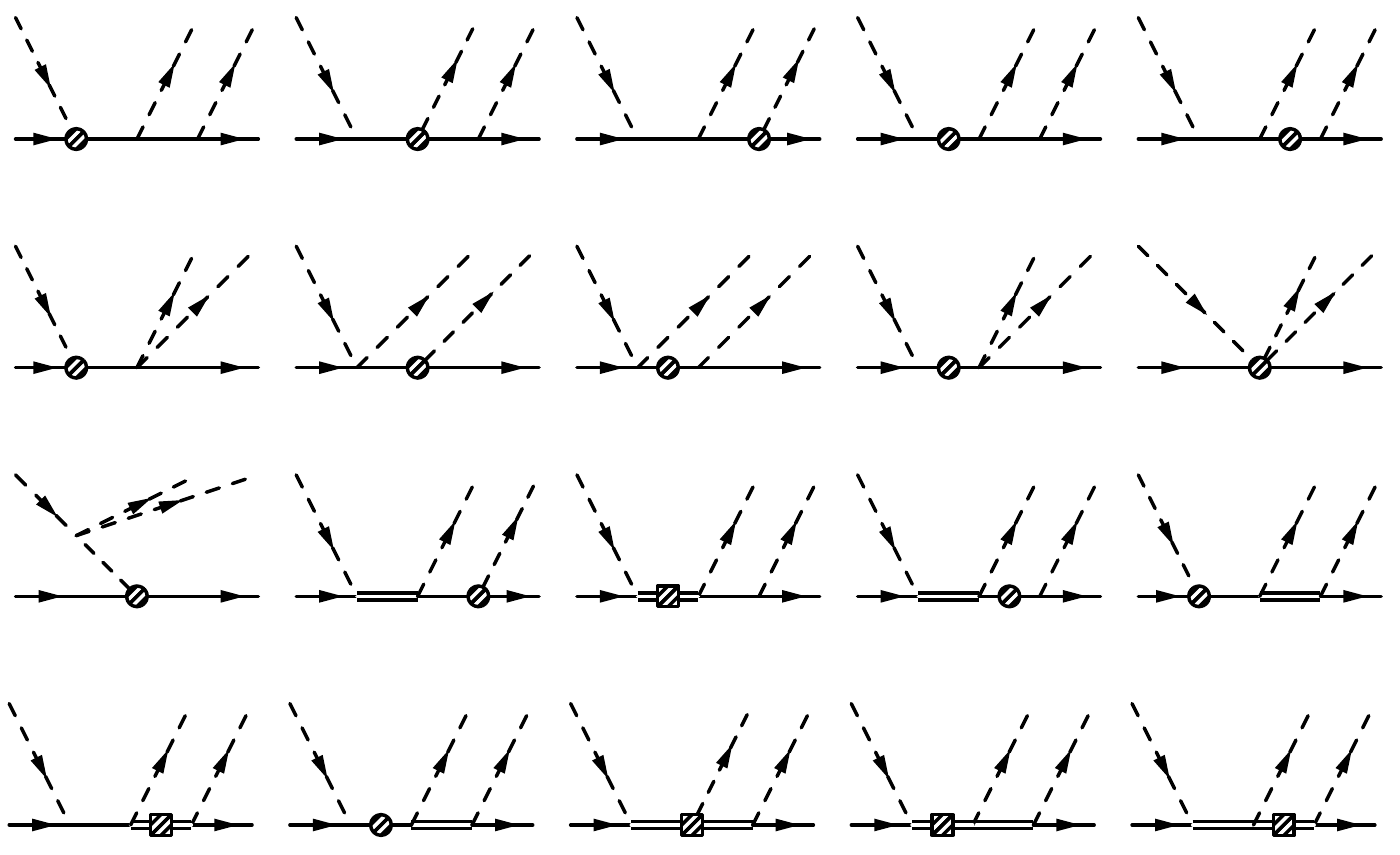}
  \caption{Additional NLO graphs contributing to the reaction $\pi N\to\pi\pi N$
    in the heavy-baryon framework. The shaded blob/square denotes
a pure $1/m_N$ insertion. Crossed
    diagrams are not shown.}
  \label{fig:NLO1Overm}
\end{figure}

When performing the calculation within the heavy-baryon framework, one
needs to take into account additional diagrams involving
$1/m_N$-vertices shown in Fig.~\ref{fig:NLO1Overm}.  
Notice that these vertices are fixed by the Poincar\'{e} invariance and do
not involve any additional parameters. 

We further emphasize that given the fact that the delta
isobar is an unstable particle, it is not appropriate to use the free delta propagator given
in Eqs.~(\ref{eq:18}) in the resonance region corresponding to $p^2 \sim
m_\Delta^2$. In particular,  a dressing of the delta becomes necessary,
resumming
all one-delta-irreducible
diagrams which obviously become large in the kinematical region with $p^2 - m_\Delta^2 =
\mathcal{O} (M_\pi^2)$, see Ref.~\cite{Pascalutsa:2002pi} for details.  
Here we will take this effect into account using the following
simple expressions for the $\Delta$ propagator where in particular the imaginary part of the derivative of the
self-energy has been neglected, see Ref.~\cite{Pascalutsa:2002pi}. In
the relativistic framework it reads
\begin{equation}
  \label{eq:18a}
\mathcal{G}^{\mu\nu}_\Delta (p)= -\frac{\slashed{p}+m_\Delta}{p^2-m_\Delta^2+\i m_\Delta \Gamma}\left(
      g^{\mu\nu}-\frac{1}{3}\gamma^\mu \gamma^\nu
      +\frac{1}{3}\frac{p^\mu \gamma^\nu-p^\nu \gamma^\mu}{m_\Delta}
      -\frac{2}{3}\frac{p^\mu p^\nu}{m_\Delta^2}\right) 
\komma
\end{equation}
with $\Gamma$ being the decay width of the $\Delta(1232)$ resonance,
while the expression in the heavy-baryon framework has a simpler form 
\begin{equation}
  \label{eq:23a}
   \mathcal{\hat{G}}_\Delta^{\mu\nu} (p)=\frac{- 1}{v\cdot
     k-\Delta+\frac{\i}{2}\Gamma}\left(P^{3/2}_{33}\right)^{\mu\nu}\xi^{ij}_{3/2}\punkt
\end{equation}
At the level of accuracy of our calculations
this should be completely sufficient. In fact, in Ref.~\cite{Jensen:1997em} the energy
dependence of the width was incorporated.
A more refined and consistent treatment
of the delta propagator will be done in a future work. 
 
While  ``dressing'' of the delta is, strictly speaking, only required in the resonance
region, we will use the expressions in Eqs.~(\ref{eq:18a})
and (\ref{eq:23a}) for the Delta propagator in all diagrams and for
all kinematical regions. Given that $\Gamma \sim \mathcal{O}
(M_\pi^3)$, such a procedure affects contributions to the
amplitude at orders $\epsilon^3$ which are beyond the scope of the
present work. 

For the nucleonic contributions to the scattering amplitude, the
results within the covariant and heavy-baryon frameworks 
we obtain agree with the ones published in Refs.~\cite{Bernard:1995gx} 
and \cite{Fettes:1999wp}. The expressions for the delta contributions
to the amplitude  are too involved to be given here but can be made
available as a Mathematica notebook upon request.

\section{Observables}
\label{sec:obs}

To match the conventions adopted by
the experimentalists, one needs to calculate the differential cross
sections with respect to different kinematical variables. In
particular, there are differential
cross sections with respect to the outgoing pion energies and solid
angles. Thus, it is advantageous to express the integration variables
in spherical coordinates and pion energies. There is a second set of
differential cross sections, which are with respect to the kinematics of
the final dipion system. These can be derived from the first set.

\begin{figure}[b]
  \centering
  \subfigure[angles $\theta_1$, $\theta_2$ and $\phi_2$]
{\includegraphics[width=0.225\textwidth]{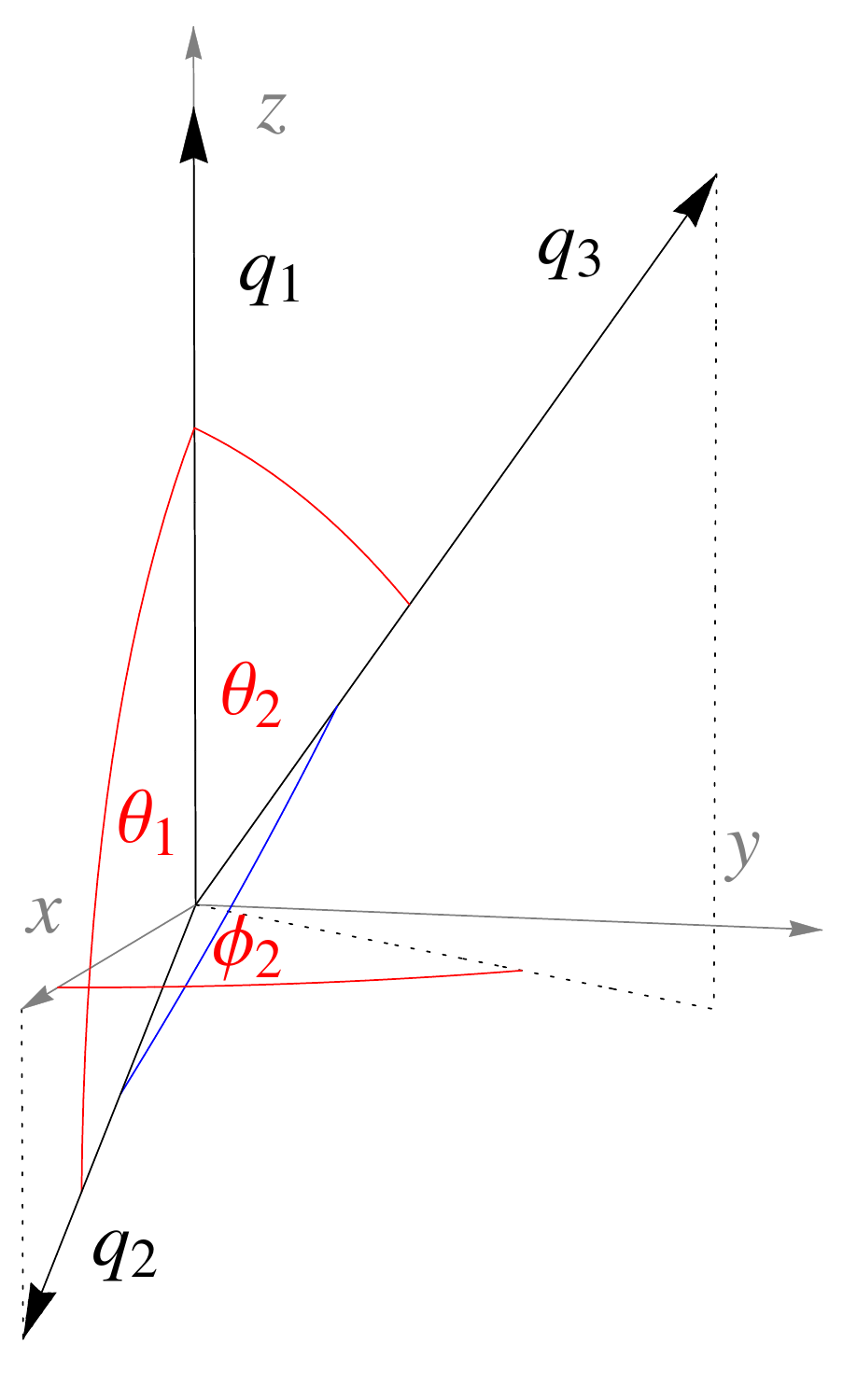}\label{fig:kinsphnz}}
  \subfigure[angle $\phi$ ]{\includegraphics[width=0.25\textwidth]{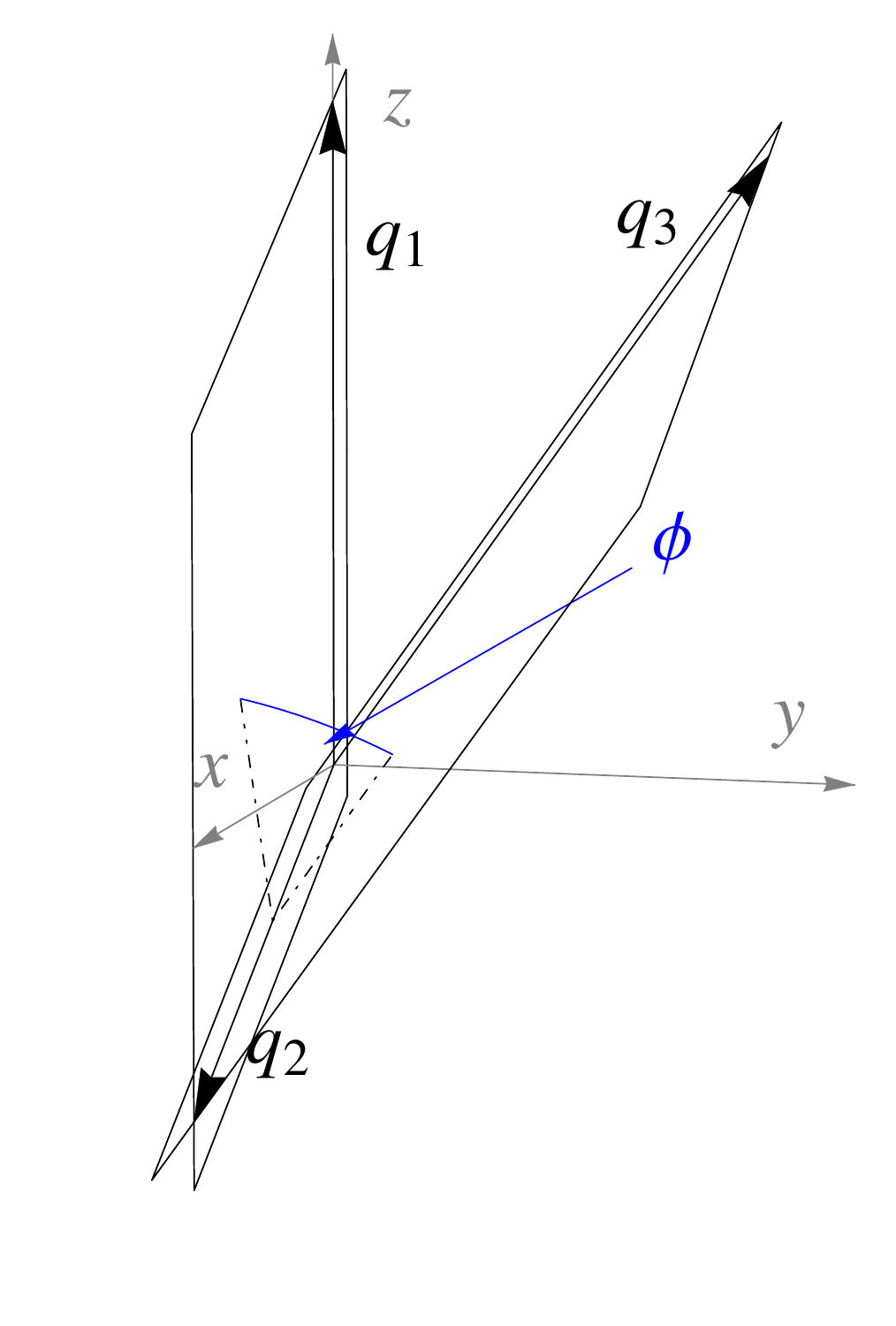}\label{fig:kinphi}}
 \subfigure[angles $\alpha$ and $\beta$]{\includegraphics[width=0.25\textwidth]{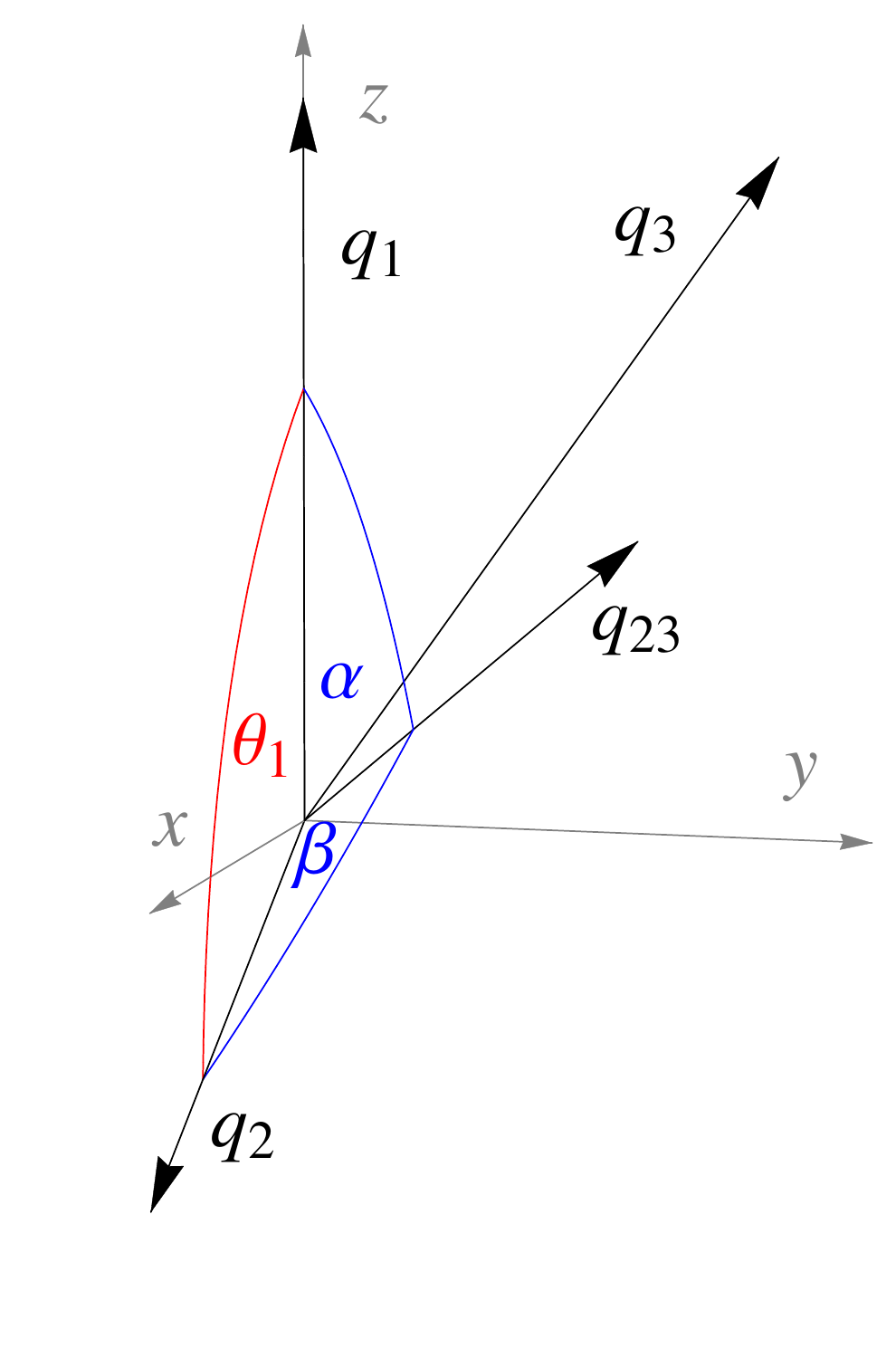}\label{fig:kinalnbe}}
  \subfigure[angle $\phi^\prime$]{\includegraphics[width=0.25\textwidth]{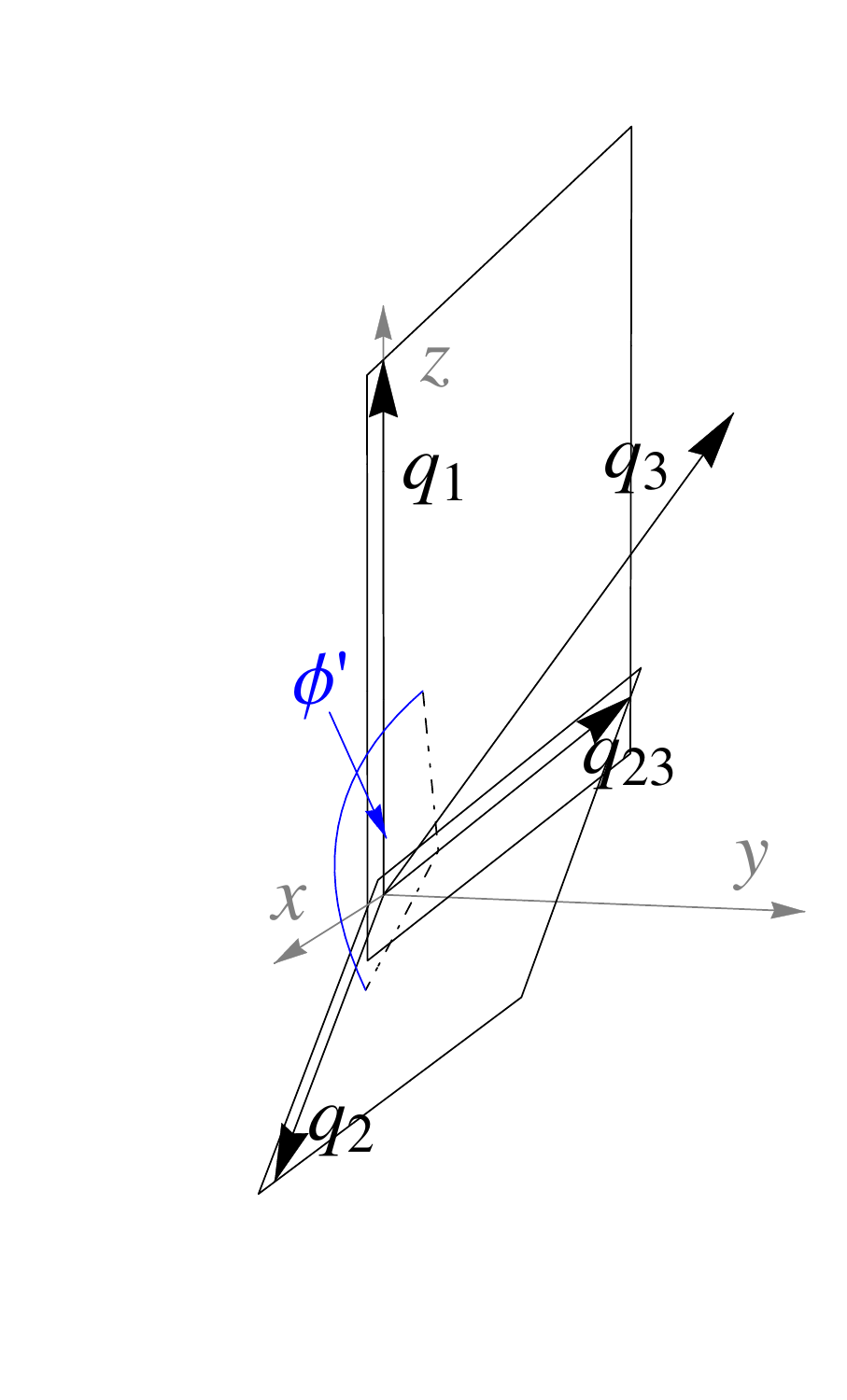}\label{fig:kinphip}}
  \caption{Kinematics and the definitions of angles} 
  \label{fig:kinematics}
\end{figure}

The experimentalists' convention suggests to choose the
coordinate frame such that $\vec{q}_1$
defines the $z$-direction and $\vec{q}_2$ lies in the $xz$-plane so that, by
construction, the azimuthal angle of $\pi^b$ is zero, see
Fig. \ref{fig:kinsphnz}. It is advantageous to introduce an auxiliary
angle $\phi$, see Fig. \ref{fig:kinphi}, which is related to the azimuthal
angles $x_i=\cos\theta_i$ via
\begin{equation}
  \label{eq:39}
  x_2 = x_1 z + \sqrt{(1-x_1^2)(1-z^2)}\cos{\phi}\punkt
\end{equation}
The formulae for the total and the double- and triple-differential
cross sections of the first set thus read
\begin{align}
  \label{eq:40}
  \begin{aligned}
    \sigma &= \frac{4 m_N^2 \mathcal{S}}{(4\pi)^4\sqrt{s}|\vec{q}_1|}
    \int_{\omega_2^-}^{\omega_2^+}{\di\omega_2
      \int_{\omega_3^-}^{\omega_3^+}{\di\omega_3\int_{-1}^1{\di
          x_1\int_0^\pi{\di \phi \;|\mathcal{M}|^2
          }} }}\komma\\
    \frac{\di^2\sigma}{\di\omega_2\di\Omega_2}&= \frac{8 m^2_N
      \mathcal{S}}{(4\pi)^5\sqrt{s}|\vec{q}_1|}
    \int_{\omega_3^-}^{\omega_3^+}{\di\omega_3\int_0^\pi{\di \phi
        \;|\mathcal{M}|^2
      }}\komma\\
    \frac{\di^3\sigma}{\di\omega_2\di\Omega_2\di\Omega_3}&= \frac{4
      m_N^2
      \mathcal{S}|\vec{q}_2||\vec{q}_3|}{(4\pi)^5\sqrt{s}|\vec{q}_1|\tilde{p}_0}
    |\mathcal{M}|^2\komma
  \end{aligned}
\end{align}
where the integration limits are given by
\begin{equation}
\begin{aligned}
  \label{eq:41}
  \omega_3^\pm = \frac{1}{2(s-2\sqrt{s}\,\omega_2+M_\pi^2)}
  \bigg[ (\sqrt{s}-\omega_2)(s-2\sqrt{s}\,\omega_2-m_N^2+2M_\pi^2) \\
  \pm |\vec{q}_2| \sqrt{(s-2\sqrt{s}\,\omega_2-m_N^2)^2-4m_N^2M_\pi^2}\bigg]
\end{aligned}
\end{equation}
and 
\begin{equation}
  \label{eq:42}
  \omega_2^-=M_\pi\quad\komma\quad
  \omega_2^+=\frac{(\sqrt{s}-M_\pi)^2-m_N^2+M_\pi^2}{2(\sqrt{s}-M_\pi)}\punkt
\end{equation}
$\mathcal{S}$ is a Bose symmetry factor: $\mathcal{S}=1/2$ for
identical outgoing pions and $\mathcal{S}=1$ otherwise. 
The kinematical variables in the center-of-mass (CMS) system are
defined according to  
\begin{align}
  \label{eq:43}
  \begin{aligned}
    s&=(m_N+M_\pi)^2+2mT_\pi\komma
    &\omega_1&=\frac{s+M_\pi^2-m_N^2}{2\sqrt{s}}\komma\\
    s_1&= s-2\sqrt{s}\, \omega_3 + M_\pi^2 \komma
    &s_2&= s-2\sqrt{s}\, \omega_2 + M_\pi^2 \komma \\
    t_1&= 2(M_\pi^2 - \omega_1\omega_2 + \vec{q}_1\cdot\vec{q}_2)
    \komma
    &t_2&= 2(M_\pi^2 - \omega_1\omega_3 + \vec{q}_1\cdot\vec{q}_3)\komma\\
    |\vec{q}_2||\vec{q}_3|z&=\omega_2 \omega_3
    -\sqrt{s}(\omega_2+\omega_3)+M_\pi^2+\frac{1}{2}(s-m_N^2)\komma
 \end{aligned}
\end{align}
where $T_\pi$
is the kinetic energy of the incoming pion in the laboratory
frame. Furthermore,
\begin{equation}
  \label{eq:44}
  \tilde{p}_0= \frac{\omega_3\left(\frac{1}{2} (s-m_N^2)-\sqrt{s}\,\omega_2\right)+ M_\pi^2(\omega_2+\omega_3-\sqrt{s}) }{\omega_3^2-M_\pi^2}\punkt
\end{equation}
Using the double- and triple-differential cross sections in
Eq. \eqref{eq:40}, one defines the angular correlation function
$W$ as follows
\begin{equation}
  \label{eq:45}
  W= 4\pi
  \left( \frac{\di^3\sigma}{\di\omega_2\di\Omega_2\di\Omega_3}\middle/
\frac{\di^2\sigma}{\di\omega_2\di\Omega_2} \right)\punkt
\end{equation}
The second set of differential cross sections is with respect to the
final dipion system with $q_{23}=q_2+q_3$, see
Fig. \ref{fig:kinalnbe}. Again, it is advantageous to define an
auxiliary angle $\phi^\prime$, see Fig. \ref{fig:kinphip}, which is related to $x_1$ via
\begin{equation}
  \label{eq:46}
  x_1 = \cos\beta \cos\alpha + \sqrt{(1-\cos^2\alpha)(1-\cos^2\beta)}\cos\phi^\prime\punkt
\end{equation}
Denoting the final dipion mass by $M_{\pi\pi}^2$, the scattering angle
of the two outgoing pions in the CMS of the final dipion by
$\theta$ and with $t=(q_1-q_2-q_3)^2$ the differential cross sections read
\begin{align}
  \label{eq:47}
    \frac{\di\sigma}{\di
      M_{\pi\pi}^2}&=\frac{m_N^2\mathcal{S}}{(4\pi)^4s|\vec{q}_{23}|
       |\vec{q}_1|^2} \int{\di t
      \int{\di\omega_2\int_0^\pi{\di\phi^\prime
          \;|\mathcal{M}|^2}}}\komma\nonumber\\
    \frac{\di\sigma}{\di t}&=
    \frac{m_N^2\mathcal{S}}{(4\pi)^4 s |\vec{q}_1|^2} \int{\frac{\di
        M_{\pi\pi}^2}{|\vec{q}_{23}|}
      \int{\di\omega_2\int_0^\pi{\di\phi^\prime
          \;|\mathcal{M}|^2}}}\komma\\
    \frac{\di\sigma}{\di M_{\pi\pi}^2\di
      t}&=\frac{m_N^2\mathcal{S}}{(4\pi)^4s|\vec{q}_{23}|
       |\vec{q}_1|^2} \int{\di\omega_2\int_0^\pi{\di\phi^\prime
        \;|\mathcal{M}|^2}}\komma\nonumber\\
    \frac{\di\sigma}{\di\cos\theta}&=
    \frac{2m_N^2\mathcal{S}}{(4\pi)^4 s |\vec{q}_1|^2} \int{\di
      M_{\pi\pi}^2 \int{\di\cos\alpha \int{\di\omega_2\;        
\frac{|\mathcal{M}|^2|\vec{q}_1^\prime||\vec{q}_2^\prime|}{|\vec{q}_2|\sin\phi^\prime
      \sqrt{(1-\cos^2\alpha)(1-\cos^2\beta)}}}}}\punkt\nonumber
\end{align}
The integration boundaries for $t$ are given by
\begin{equation}
  \label{eq:48}
  t^\pm=M_\pi^2+M_{\pi\pi}^2-2\omega_1\omega_{23}\pm
2|\vec{q}_1||\vec{q}_{23}|
\qquad\mathrm{with}\qquad
  \omega_{23}=\omega_2+\omega_3=\frac{M_{\pi\pi}^2+s-m_N^2}{2\sqrt{s}}\punkt
\end{equation}
The integration boundaries for $\omega_2$ are restricted to the
overlap of the interval $[\omega_2^-,\omega_2^+]$ from
Eq. \eqref{eq:42} and the interval
$[\tilde{\omega}_2^-,\tilde{\omega}_2^+]$  with
\begin{equation}
  \label{eq:58}
\tilde{\omega}_2^\pm=
\frac{\omega_{23}}{2}\pm\sqrt{\frac{(\omega^2_{23}-M_{\pi\pi}^2)(M_{\pi\pi}^2-4M_{\pi}^2)}{4M_{\pi\pi}^2}}\punkt
\end{equation}
The integration boundaries for $M_{\pi\pi}^2$ are
\begin{equation}
  \label{eq:59}
  {M_{\pi\pi}^2}^\pm=\frac{1}{m_N^2}\left( \frac{t}{2}(s+m_N^2-M_\pi^2)+
    m_N^2M_\pi^2\pm |\vec{q}_1|\sqrt{-s\,t(4m_N^2-t)} \right)
\end{equation}
and also $M_{\pi\pi}^2\in(4M_\pi^2, (\sqrt{s}-m_N)^2)$. Finally, the
integration for the differential cross section with respect to the
scattering angle is restricted by the condition
$\sin\phi^\prime<1$. Furthermore, the magnitudes of the pion momenta
in the dipion CMS are
\begin{equation}
  \label{eq:60} |\vec{q}_1^\prime|^2=\frac{M_{\pi\pi}^4-2M_{\pi\pi}^2(t+M_\pi^2)+(t-M_\pi^2)^2}{4M_{\pi\pi}^2}
\komma \qquad
  |\vec{q}_2^\prime|^2=\frac{M_{\pi\pi}^2-4M_\pi^2}{4}\punkt
\end{equation}

\section{Fitting Procedure}
\label{sec:fits}

The scattering amplitude for the reaction $\pi N\to\pi\pi N$ depends on several
LECs as explained in section \ref{sec:graphs}. Throughout this work, we use the following
values for the various LECs and masses entering the leading order
effective Lagrangian: $M_\pi = 139.57$ MeV, $F_\pi =92.4$ MeV, $m_N=
938.27$ MeV, $g_A=1.26$, $\Delta = 294$ MeV, $\Gamma = 118$ MeV. For
the $\pi N \Delta$ axial coupling constant $h_A$, we adopt the same
value as used in Ref.~\cite{Krebs:2013}, namely $h_A = 1.34$,
corresponding to the large-$N_c$ prediction. Notice that this value is
close to the one determined from the width  of the delta resonance (in
the covariant framework).  At  next-to-leading
order we encounter contributions proportional to the LECs $c_i$ from
$\mathcal{L}_{\pi N}^{(2)}$. Since we intend to investigate, among
others, the role played by the $\Delta$ isobar by comparing the
results obtained within the deltaless and deltafull formulations of
chiral EFT, we adopt here the values of the $c_i$'s determined from 
the fits to pion-nucleon scattering of
Refs.~\cite{Krebs:2012yv,Krebs:2013} and collected in Table \ref{tab:LECc}. 
\begin{table}[tb]
  \centering
\subtable[deltafull $\chi$PT]{
  \begin{tabular}{|c| c | c | c | c |}
\hline
    &$c_1$ &$c_2$ &$c_3$  &$c_4$ \\\hline \hline
KH &-0.95 &1.90 &-1.78 &1.50\\\hline
GW &-1.41 &1.84 &-2.55 &1.87\\\hline
  \end{tabular}}\hskip 20pt
\subtable[deltaless $\chi$PT]{
  \begin{tabular}{|c| c | c | c | c |}
\hline
    &$c_1$ &$c_2$ &$c_3$  &$c_4$ \\\hline \hline
KH &-0.75 &3.49 &-4.77 &3.34\\\hline
GW &-1.13 &3.69 &-5.51 &3.71\\\hline
  \end{tabular}}
\caption{LECs $c_i$ from the pion-nucleon sector for two different fits in
  a deltafull and deltaless theory. All values are given in GeV$^{-1}$.}
\label{tab:LECc}
\end{table}
These fits have been performed to the partial wave analyses of the 
group at the George Washington University \cite{Arndt:2006bf} (GW) and the
Karlsruhe-Helsinki analysis \cite{Koch:1985bn} (KH) at the subleading
one-loop order of HB$\chi$PT with and without explicit $\Delta$ and
using the same values of various parameters as specified above.  
The values for the deltaless approach are consistent with the 
theoretically cleaner determination
from inside the Mandelstam triangle \cite{Buettiker:1999ap}.
Thus, given that the values of the LECs $c_i$ are fixed, there are no
free parameter in the deltaless approach at order $q^2$.  

Further LECs contribute to the amplitude in the deltafull approach at 
order  $\varepsilon^2$. In particular, in addition to the $c_i$'s from $\mathcal{L}_{\pi N}^{(2)}$, 
there are also contribution  involving the LECs $b_3$,
$b_4$, $b_5$, $b_6$
from $\mathcal{L}_{\pi N \Delta}^{(2)}$ and $g_1$ from
$\mathcal{L}_{\pi \Delta}^{(1)}$ whose determination will be discussed
below. Notice that there is a large-$N_c$ prediction for the
leading-order pion-delta coupling constant $g_1$, namely
\begin{equation}
 g_1=\frac{9}{5}g_A= 2.27\komma
\label{eq:106}
\end{equation}
which will be used in some fits as will be described below. 

Isospin breaking is accounted for in a minimal way by shifting
$T_\pi$ from the isospin symmetric threshold to the physical threshold
of each channel \cite{Bernard:1997tq}. 
In the lab frame, the incoming pion kinetic energy at
threshold is
\begin{equation}
  \label{eq:87}
  T_\pi^{\mathrm{thr}}
=\frac{(M_{\pi^b}+M_{\pi^c}+m^\prime_N)^2-(m_N+M_{\pi^a})^2}{2m_N}\komma
\end{equation}
whereas the isospin symmetric case ($m^\prime_N=m_N$, $M_{\pi^a}=M_{\pi^b}=M_{\pi^c}=M_\pi$) yields
\begin{equation}
  \label{eq:88}
  T_\pi^{\mathrm{thr,iso}}=M_\pi \left( 1+\frac{3M_\pi}{2m_N} \right)=\unit{170.71}{\mega\electronvolt}\punkt
\end{equation}
The resulting shifts $\delta T_\pi = T_\pi^{\mathrm{thr}}-
T_\pi^{\mathrm{thr,iso}}$ for each channel are thus
\begin{align}
\label{eq:97}
  \begin{aligned}
    &\mathrm{I.} &\pi^-p&\to \pi^0\pi^0n:   &\delta T_\pi &= \unit{-10.21}{\mega\electronvolt} &\\
    &\mathrm{II.} &\pi^-p&\to \pi^+\pi^-n: &\delta T_\pi &=
    \unit{+1.68}{\mega\electronvolt}\\
    &\mathrm{III.} &\pi^+p&\to \pi^+\pi^+n: &\delta T_\pi &= \unit{+1.68}{\mega\electronvolt}\\
    &\mathrm{IV.} &\pi^+p&\to \pi^+\pi^0p:  &\delta T_\pi &=
    \unit{-5.95}{\mega\electronvolt}\\
    &\mathrm{V.} &\pi^-p&\to \pi^0\pi^-p:  &\delta T_\pi &= \unit{-5.95}{\mega\electronvolt}\punkt
  \end{aligned}
\end{align}

All fits described below are performed globally to the experimental
data in all five channels simultaneously. For the
fitting procedure, only the total cross section data were used, which are
taken from the compilation \cite{Vereshagin:1995mm} and
from \cite{Kermani:1998gp} and \cite{Lange:1998ti}. 
The $\chi^2$ is given by the square of the
difference between our calculated values of the total cross section and the experimental central
values divided by the squared errors on the latter. Typically a very good fit
corresponds to a $\chi^2$ per degree of freedom, $\chi^2/\mathrm{dof}$,  close to one.

\subsection{Heavy-baryon chiral perturbation theory}
Given the expected validity range of HB$\chi$PT, only data
with $T_\pi < \unit{250}{\mega\electronvolt}$ were used in the
fits. The choice of this still rather large energy is motivated by the fact
that in some channels there is essentially no data in the very near
threshold region.
 
In the static limit, the LECs $b_3$ and $b_6$ are
redundant because they can be fully absorbed into shifts of other LECs \cite{Long:2010kt}
\begin{equation}
  \label{eq:89}
  \begin{aligned}
    h_{A}&\to h_{A} -\Delta\, (b_3+b_6) \komma\\
    c_2&\to c_2 + \frac{8}{9}h_{A}\,(b_3+b_6) \komma\\
    c_3 &\to c_3 - \frac{8}{9}h_{A}\, (b_3+b_6) \komma\\
    c_4 &\to c_4 + \frac{4}{9}h_{A}\, (b_3+b_6) \komma\\
    b_4 &\to b_4 +\left(\frac{13}{9}g_1-g_A\right) (b_3+b_6) \komma\\
    b_5 &\to b_5 -\frac{4}{3}g_1 \,(b_3+b_6)\komma
  \end{aligned}
\end{equation}
which is equivalent to setting $b_3=-b_6$ in the amplitudes as done in
Ref.~\cite{Krebs:2013}. Thus, the only free parameters one is left
with at NLO are $g_1$, $b_4$ and $b_5$.

\subsection{Relativistic chiral perturbation theory}
In this case, we have carried out the fits using the same energy
interval as in the HB approach,
$T_\pi<\unit{250}{\mega\electronvolt}$, as well as  a larger energy
range of $T_\pi<\unit{400}{\mega\electronvolt}$.  
Notice that contrary to the HB formulation, the LECs  $b_3$ and $b_6$
are no more redundant 
at order $\varepsilon^2$ and have to be determined from the data, so
that one is left with a total of five unknown parameters. Furthermore,
given that we employ here the values of the LECs $c_i$ of Ref.~
\cite{Krebs:2013} as input in our analysis and in order to ensure a meaningful
comparison with the HB$\chi$PT results, we have to account for the shifts
induced in the LECs $h_A$, $c_{2,3,4}$ and $b_{4,5}$ by the nonvanishing 
linear combination $b_3 + b_6$ as specified in Eq.~(\ref{eq:89}).

\subsection{Results}
\begin{figure}[ht]
  \centering
 \includegraphics[width=0.45\textwidth]{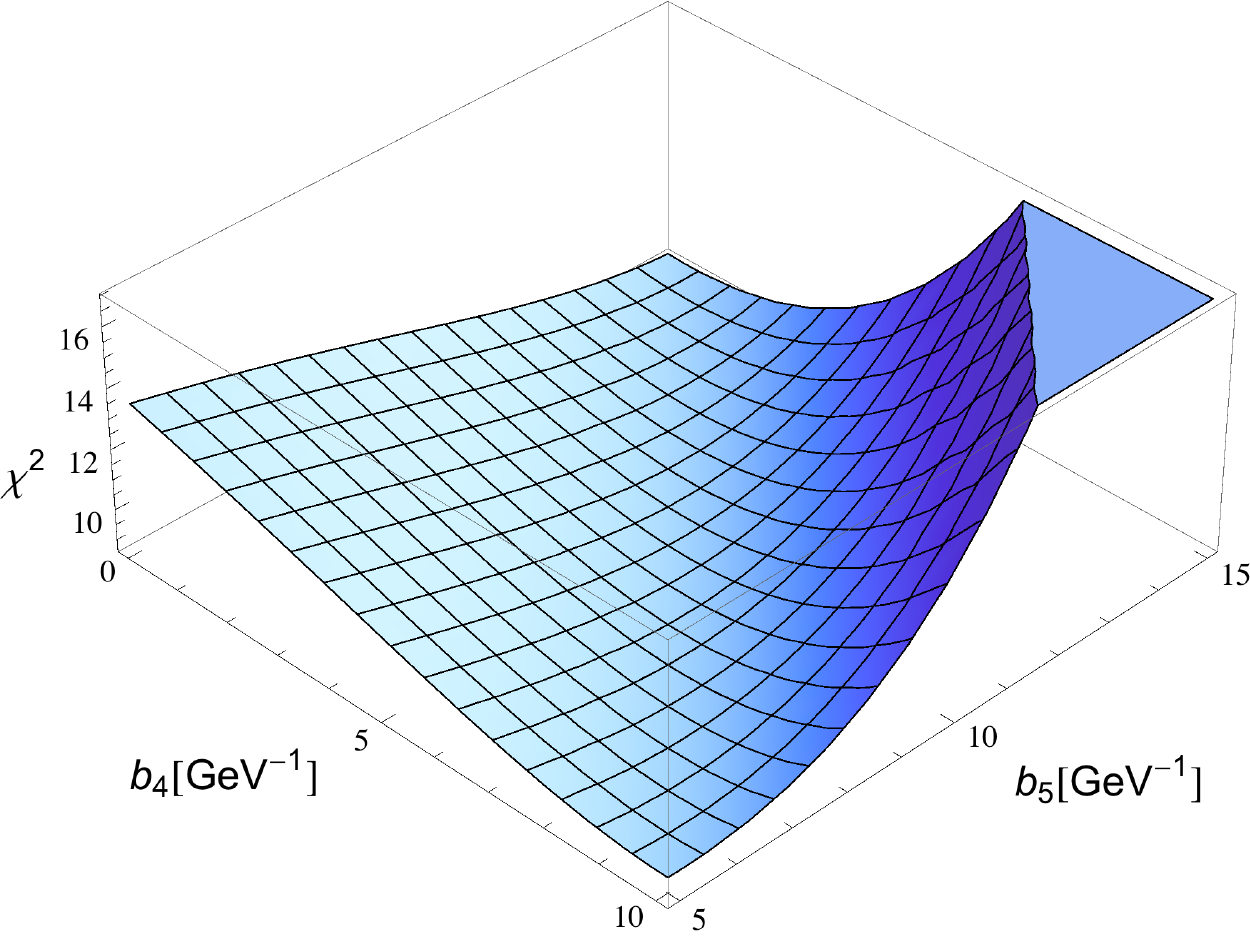} 
  \caption{Fit in HB$\chi$PT: Anticorrelation between $b_4$ and $b_5$ ($g_1=2.27$).}
  \label{fig:anticorrb4b5HB}
\end{figure}

\begin{figure}[ht]
\vskip -7 true cm
  \centering
\subfigure[$b_3=0.99$ GeV$^{-1}$, $b_6=-0.61$ GeV$^{-1}$]{
 \includegraphics[width=0.45\textwidth]{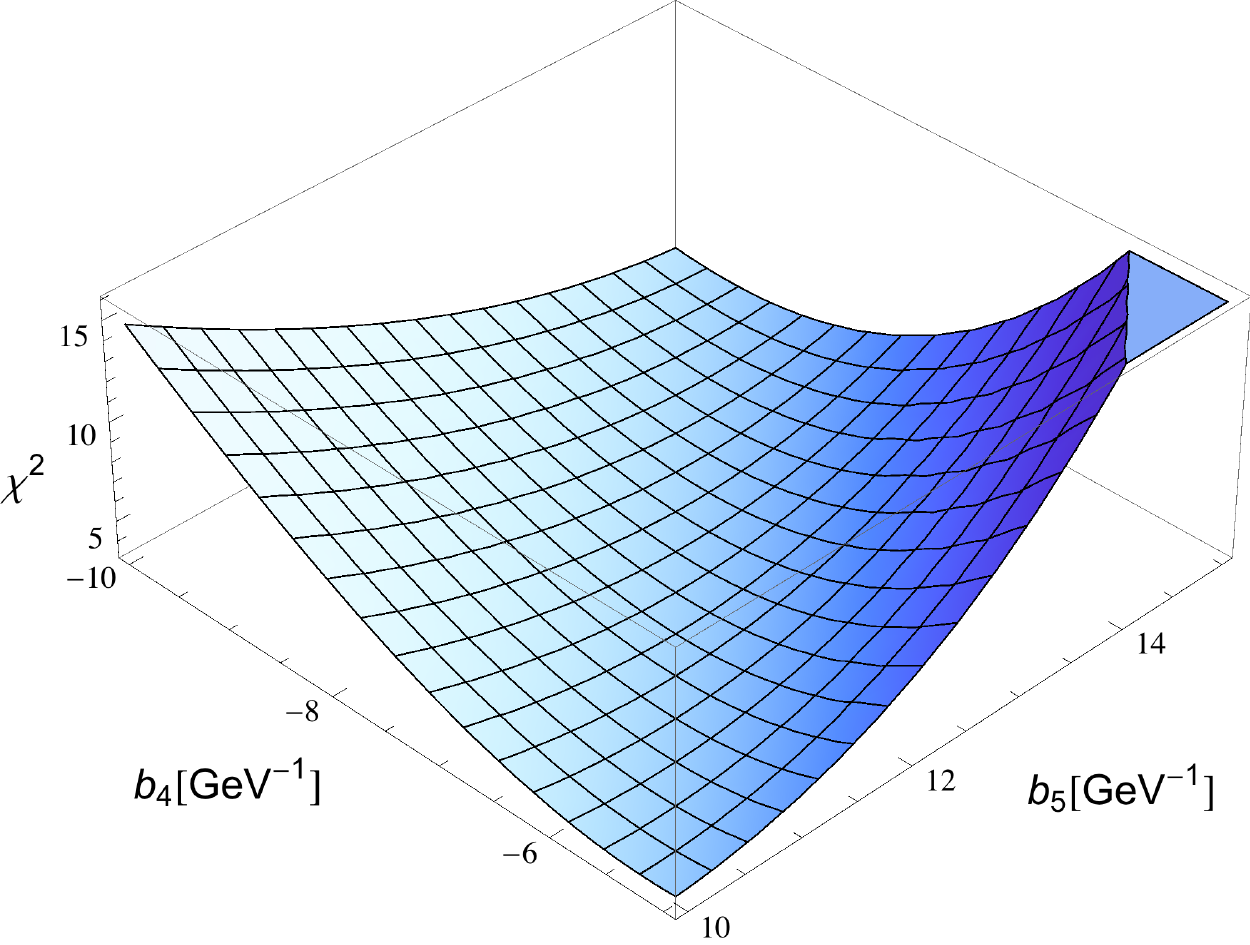}}
\subfigure[$b_4=-6.37$ GeV$^{-1}$, $b_5=11.34$ GeV$^{-1}$]{
 \includegraphics[width=0.45\textwidth]{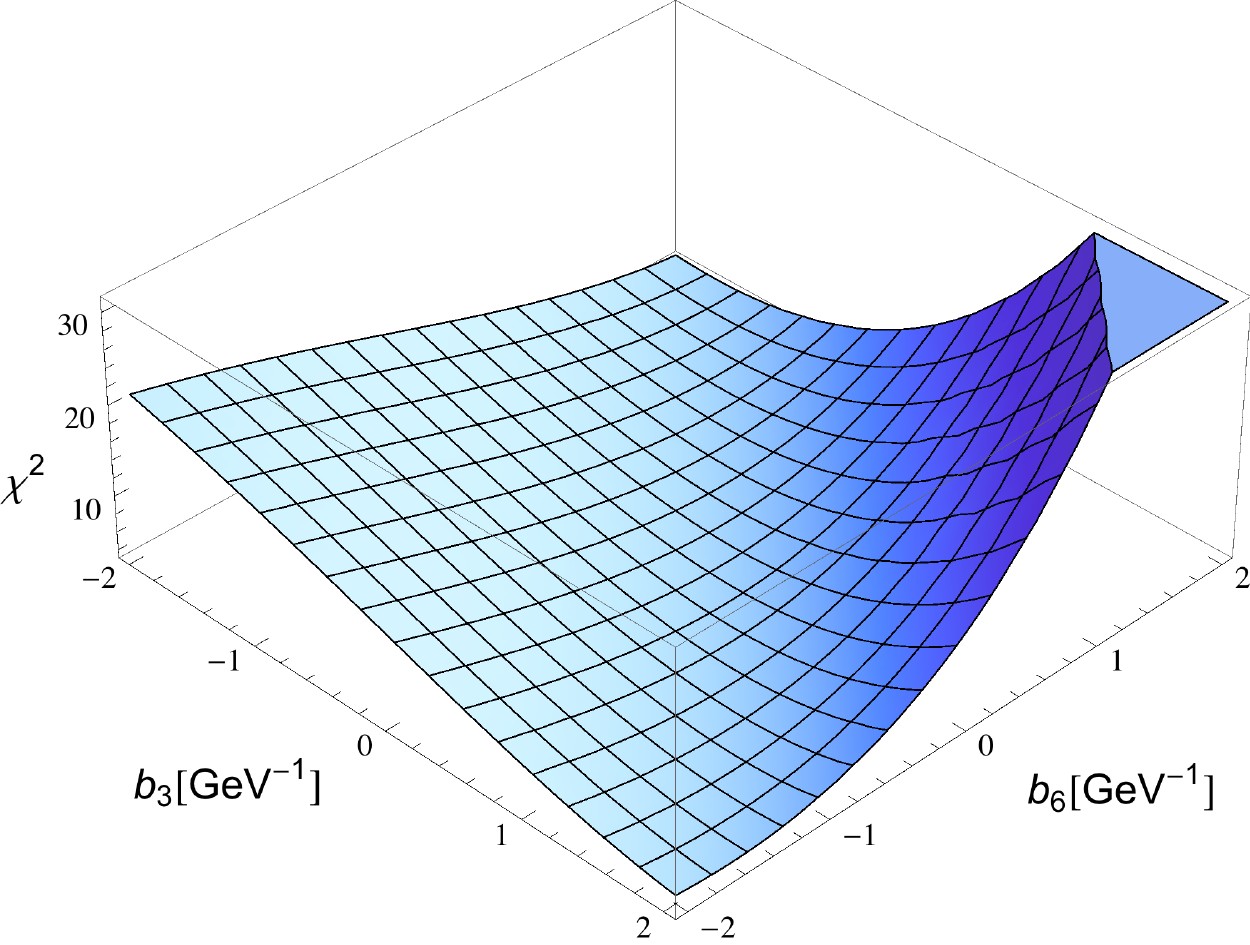}}
  \caption{Fit in $\chi$PT: Anticorrelation between $b_4$ and $b_5$
    and between $b_3$ and $b_6$
    ($g_1=2.27$).}
  \label{fig:anticorrRel}
\end{figure}

At NLO several fits were performed, both with a free and fixed value
of $g_1$. The results with the $c_i$'s taken from the upper row of Table \ref{tab:LECc}(a)
(KH) are summarized in Table \ref{tab:KHcFit}. 
\begin{table}[h]
  \centering
  \begin{tabular}{|c|rcl|rcl|rcl|rcl|rcl|c|}\hline
    Fit with KH $c_i$&\multicolumn{3}{c|}{$g_1$} &\multicolumn{3}{c|}{$b_4+b_5$}
    &\multicolumn{3}{c|}{$b_4-b_5$}
    &\multicolumn{3}{c|}{$b_3+b_6$} &\multicolumn{3}{c|}{$b_3-b_6$} &$\chi^2/\mathrm{dof}$\\\hline\hline
   \multirow{2}{*}{HB: $T_\pi<\unit{250}{\mega\electronvolt}$}
&$1.36$&$\pm$&$0.73$ &$16.61$&$\pm$&$0.66$ &$-1.75$&$\pm$&$9.78$
&\multicolumn{3}{c|}{------} &\multicolumn{3}{c|}{------} &9.42\\ 
&\multicolumn{3}{c|}{$2.27^*$} &$16.00$&$\pm$&$0.37$
&$-7.99$&$\pm$&$5.71$ &\multicolumn{3}{c|}{------} &\multicolumn{3}{c|}{------} &9.63\\\hline
   \multirow{2}{*}{Rel: $T_\pi<\unit{250}{\mega\electronvolt}$}
&$1.68$&$\pm$&$1.38$ &$5.05$&$\pm$&$0.39$ &$-12.08$&$\pm$&$15.59$ &$0.08$&$\pm$&$1.39$
&$-1.02$&$\pm$&$8.40$ &3.36\\ 
&\multicolumn{3}{c|}{$2.27^*$} &$4.97$&$\pm$&$0.29$ &$-17.71$&$\pm$&$12.22$ &$0.39$&$\pm$&$0.86$
&$1.60$&$\pm$&$7.62$ &3.40\\\hline
   \multirow{2}{*}{Rel: $T_\pi<\unit{400}{\mega\electronvolt}$}
&$1.41$&$\pm$&$0.22$ &$4.35$&$\pm$&$0.08$ &$1.62$&$\pm$&$1.87$ &$-1.07$&$\pm$&$0.18$
&$3.14$&$\pm$&$1.61$ &4.26\\ 
&\multicolumn{3}{c|}{$2.27^*$} &$4.51$&$\pm$&$0.12$ &$1.17$&$\pm$&$1.61$
&$-1.17$&$\pm$&$0.12$ &$7.36$&$\pm$&$2.03$&4.66\\\hline
  \end{tabular}
  \caption{LECs determined from global fits to the total cross section
    data at NLO using
    the KH set of LECs $c_i$  from
    Table \ref{tab:LECc}(a) as input. The star indicates that the corresponding
  value is kept fixed. The values of LECs $b_i$ are given in units
  of GeV$^{-1}$.}
\label{tab:KHcFit}
\end{table}
In all cases, we find that the LECs $b_4$ and $b_5$ are strongly
anticorrelated. This is visualized in Fig.~\ref{fig:anticorrb4b5HB}
for the heavy-baryon approach and in the left panel of
Fig.~\ref{fig:anticorrRel} for the relativistic calculation. As a
result, while the value of the linear combination $b_4+b_5$ can be
reliably extracted in each fit, there is a very large uncertainty of about
$100\%$ for the linear combination $b_4-b_5$. Further, the LECs $b_3$
and $b_6$ also appear to be strongly anticorrelated in the
relativistic approach
as shown in the right panel of
Fig.~\ref{fig:anticorrRel}.

The significance of relativistic effects in the reaction $\pi N
\to \pi \pi N$ is clearly seen in the strong reduction of $\chi^2/$dof
from $\sim 9.5$ in the HB approach to $\sim 3.4$ in relativistic
$\chi$PT.  In addition, the unnaturally large value of the linear
combination  $b_4 + b_5$ ,  $b_4 + b_5 \sim 16$
GeV$^{-1}$, in HB$\chi$PT
indicates that the energies up to $T_{\pi} = 250$ MeV employed in the
fit are probably beyond the
applicability range of the order-$\epsilon^2$ HB approximation. 
In contrast, the fits carried out within the relativistic $\chi$PT
framework lead to reasonably natural values for $b_4 + b_5$. 

In order to test the stability of our results and to get further
insights into the applicability range of relativistic $\chi$PT, we
have extended the energy range used in the fits up to $T_\pi = 400$
MeV, see the last row in Table \ref{tab:KHcFit}.  Remarkably,
including the higher-energy data only increased the value of 
$\chi^2/$dof by about $30\%$. As expected, including higher energies
in the fit stabilizes the results for the LECs which manifests itself
in the significantly reduced error bars. It is comforting to see that
the 
values of all LECs extracted from  the unconstrained fits up to  $T_\pi = 250$
MeV and  $T_\pi = 400$ MeV agree with each other within the error
bars. One also observes that the anticorrelations between the LECs 
$b_4$ and $b_5$ as well as $b_3$ and $b_6$ are much less pronounced 
in the higher-energy fit.  The situation is similar in the constrained
fits, although the deviations between the extracted LECs tend to be
somewhat  larger.      

We further observe that there is a fairly minor sensitivity to the
LEC $g_1$. In particular,  fixing $g_1$ to its
large-$N_C$ value appears to only mildly affect the $\chi^2/$dof and
also has little impact on the extracted values of other
LECs. This manifests itself in rather large error bars for $g_1$ when performing
fits up to $T_\pi = 250$ MeV. The extracted values in the HB
approach, $g_1 = 1.36 \pm 0.73$, and relativistic $\chi$PT, 
$g_1=1.68 \pm 1.38$, agree well with each other as well as with the
large-$N_c$  value of $g_1= 2.27$. The higher-energy fit within the
relativistic framework yields a similar result but with reduced error
bars, $g_1 = 1.41 \pm 0.22$, which is somewhat smaller than the 
large-$N_c$ prediction. It should, however, be emphasized that 
it is not completely consistent to fit $g_1$ while, at the same time, 
using the values for the LECs $c_1$ as input, which have been 
extracted from $\pi N$ scattering in Ref.~\cite{Krebs:2013} using  $g_1$
fixed to  its large-$N_c$ value. This could be improved in the future 
by carrying out a simultaneous analysis of the $\pi N \to \pi N$ and 
$\pi N \to \pi \pi N$ reactions.  

Last but not least, we have also carried out fits using the GW set of the
$c_i$'s from Table \ref{tab:LECc}  which lead to slightly different values for the fitted
parameters without affecting any of the conclusions. The LECs
resulting from the constrained fits up to $T_\pi = 250$ MeV  using the KH
and GW sets of LECs $c_i$ as input are listed in Table
\ref{tab:globalFit}.  
\begin{table}[tb]
  \centering
  \begin{tabular}{|c|c|rcl|rcl|rcl|rcl|rcl|c|}\hline
    Fit &$c_i$ &\multicolumn{3}{c|}{$g_1$} &\multicolumn{3}{c|}{$b_4+b_5$}
    &\multicolumn{3}{c|}{$b_4-b_5$}
    &\multicolumn{3}{c|}{$b_3+b_6$} &\multicolumn{3}{c|}{$b_3-b_6$} &$\chi^2/\mathrm{dof}$\\\hline\hline
\multirow{2}{*}{HB: $T_\pi<\unit{250}{\mega\electronvolt}$} &KH
&\multicolumn{3}{c|}{$2.27^*$} &$16.00$&$\pm$&$0.37$
&$-7.99$&$\pm$&$5.72$ &\multicolumn{3}{c|}{------} &\multicolumn{3}{c|}{------} &9.63\\\cline{2-18}
&GW 
&\multicolumn{3}{c|}{$2.27^*$} &$15.99$&$\pm$&$0.37$
&$-8.42$&$\pm$&$5.77$ &\multicolumn{3}{c|}{------} &\multicolumn{3}{c|}{------} &9.65\\\hline
   \multirow{2}{*}{Rel: $T_\pi<\unit{250}{\mega\electronvolt}$}&KH
&\multicolumn{3}{c|}{$2.27^*$} &$4.97$&$\pm$&$0.29$ &$-17.71$&$\pm$&$12.22$ &$0.39$&$\pm$&$0.86$
&$1.60$&$\pm$&$7.62$ &3.40
\\\cline{2-18}
&GW &\multicolumn{3}{c|}{$2.27^*$} &$4.34$&$\pm$&$0.29$ &$-18.24$&$\pm$&$10.77$ &$0.70$&$\pm$&$0.76$
&$1.41$&$\pm$&$6.38$ &3.47\\\hline
  \end{tabular}
  \caption{LECs determined from global fits to the total cross section
    data at NLO using the KH and GW sets of LECs $c_i$  from
    Table \ref{tab:LECc} as input. The star indicates that the corresponding
  value is kept fixed. The values of LECs $b_i$ are given in unuts
  of GeV$^{-1}$.}
\label{tab:globalFit}
\end{table}

\section{Predictions}
\label{sec:predictions}

We are now in the position to make predictions for various
observables. Here and in what follows, we will use the values of the 
LECs collected  in Table \ref{tab:globalFit}. This will
allow us to make a meaningful comparison between the predictions
in the HB and relativistic $\chi$PT.  
We also use both the KH and GW sets of the LECs $c_i$ from Table
\ref{tab:LECc} in order to estimate the uncertainty associated with
the pion-nucleon system which provides input for our
calculations. Thus, all predictions at NLO ($Q^2$ or $\epsilon^2$) 
are visualized by bands whose width corresponds to the variation of
the LECs $c_i$ between the KH and GW values. Further, while we have
analyzed all available low-energy observables in this reaction, we
only show in the following selected representative examples. 

\begin{figure}[ht]
\vspace{-1cm}
  \centering
  \includegraphics[width=0.9\textwidth]{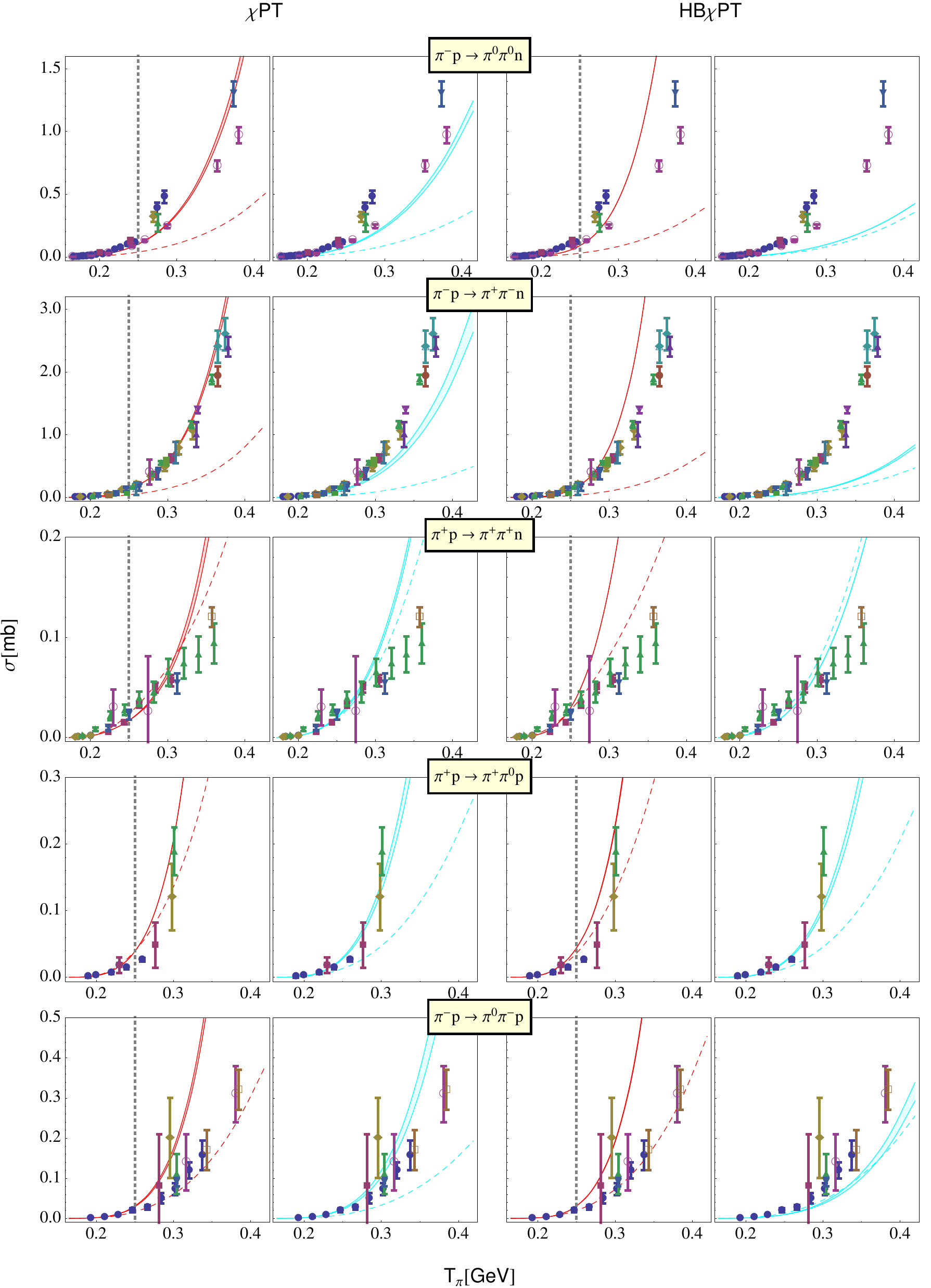}
\caption{Predictions for the total cross section up to
  $T_\pi\simeq\unit{400}{\mega\electronvolt}$. The columns from left to
  right correspond to the deltafull covariant $\chi$PT, deltaless
  covariant $\chi$PT, deltafull HB$\chi$PT and deltaless HB$\chi$PT
  predictions, respectively.     The dashed  and solid
  lines refer to LO (i.e. order $Q^1$ or $\epsilon^1$) and NLO
  (i.e.~up to order $Q^2$ or $\varepsilon^2$) results. The  energies
  used in the fit at NLO are below the vertical dotted line. The bands at
NLO correspond to using the KH and GW sets of LECs $c_i$. }
\label{fig:sigmatot}
\end{figure}

The predictions for the total cross sections with incoming pion
energies up to $\unit{400}{\mega\electronvolt}$ are presented in 
Fig.~\ref{fig:sigmatot}, both in the deltafull and deltaless
theories. 
The deltaless calculations were performed with the $c_i$'s taken
from Table \ref{tab:LECc}(b). As can be seen, the relativistic
approach describes the data at higher energies much better than the
heavy-baryon one, which is fully in line with the observations 
made in the previous section.  The
predictions within HB$\chi$PT at NLO appear to significantly underestimate the data in almost
every channel, whereas the deltafull HB predictions overshoot the
cross sections for $T_\pi>\unit{300}{\mega\electronvolt}$. The inclusion of the delta
in the relativistic case is mainly noticeable in the upper two
channels, whereas the description of the other three channels is
similar. 

\begin{figure}[ht]
  \centering
\subfigure{
  \includegraphics[width=0.78\textwidth]{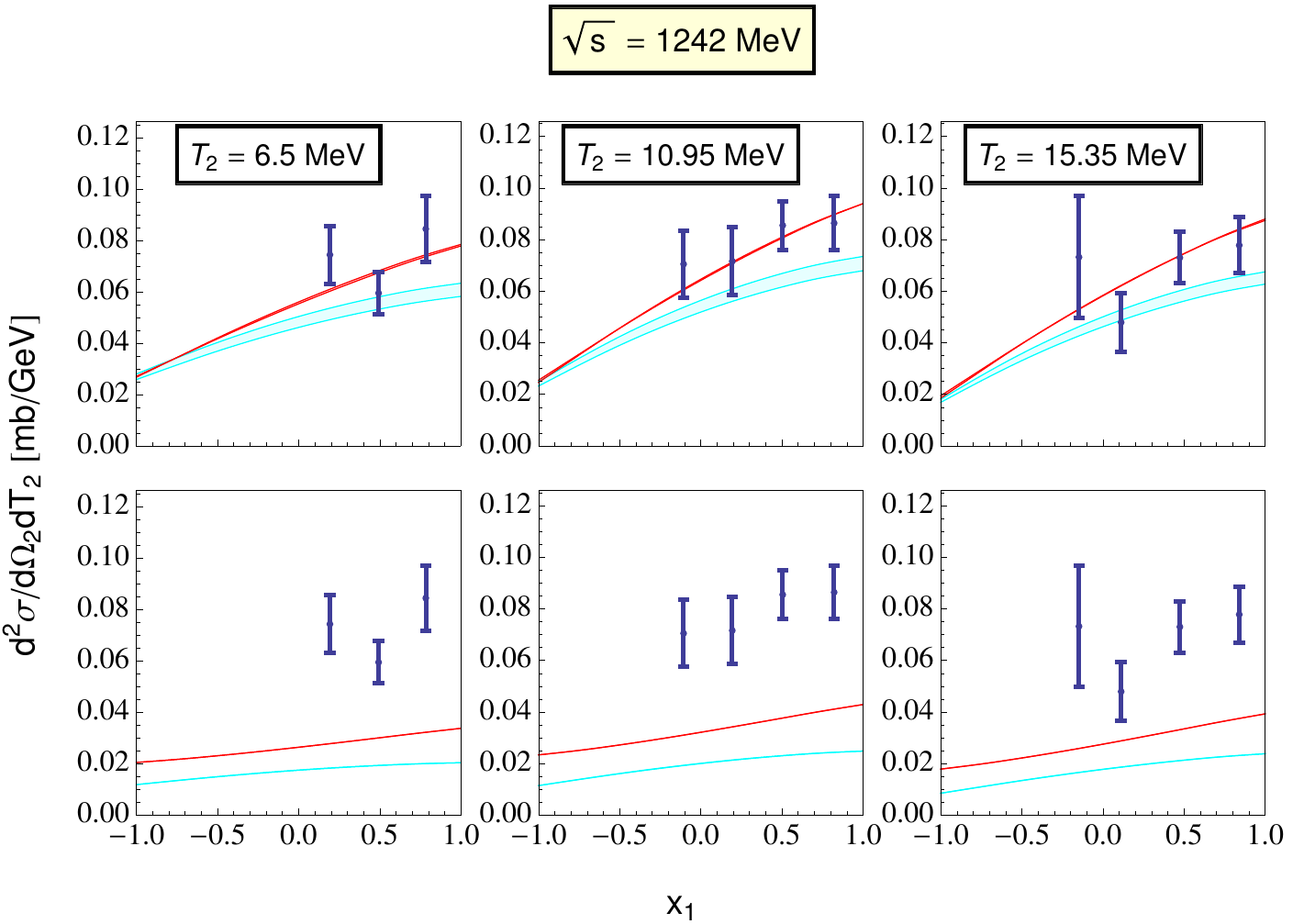}}
\subfigure{
  \includegraphics[width=1\textwidth]{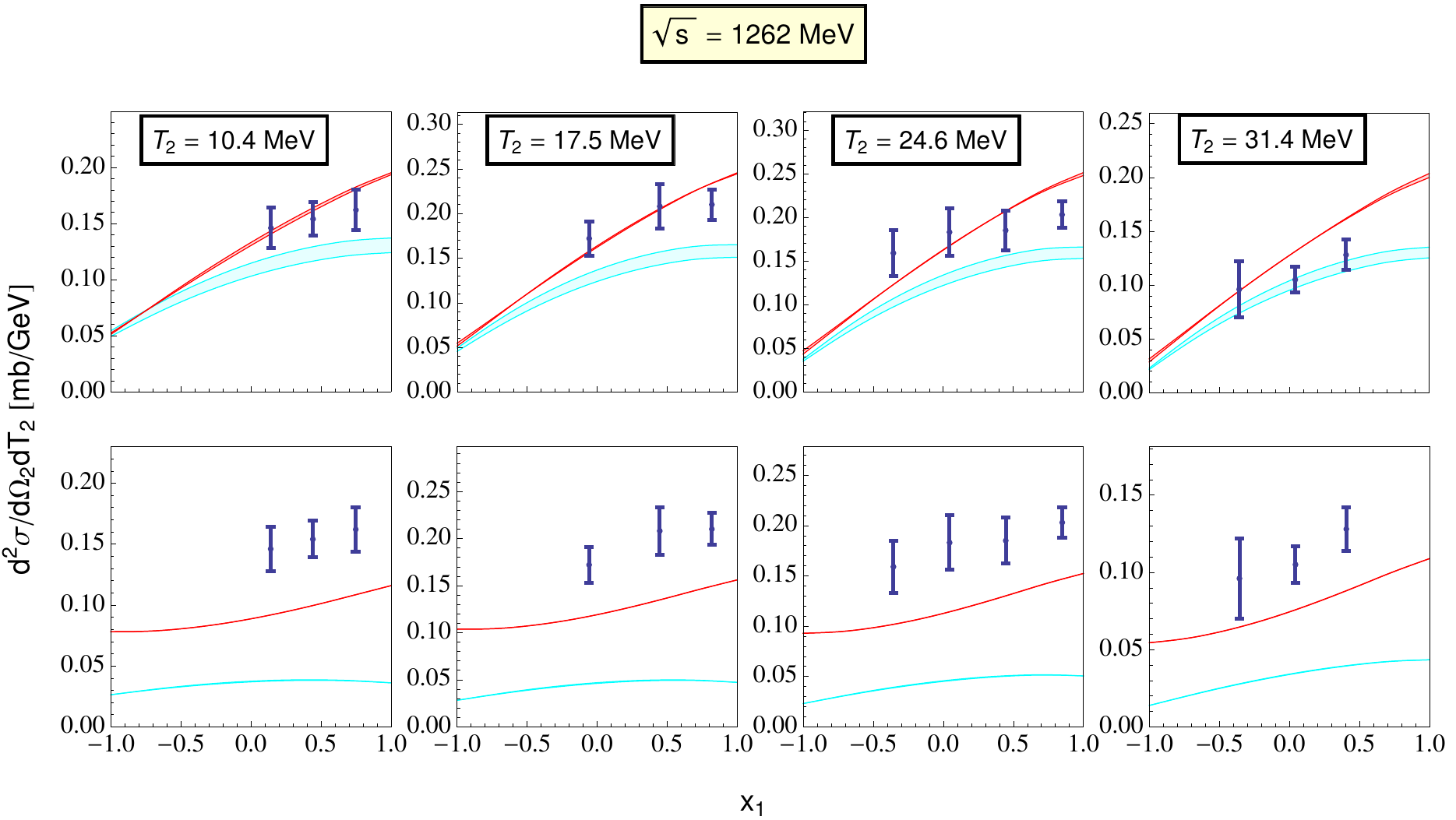}}
\caption{NLO $\chi$PT predictions for the double-differential cross sections
  for $\pi^-p\to\pi^+\pi^-n$
  with respect to the kinetic energy and the solid angle of the
  outgoing $\pi^+$, see Eq. \eqref{eq:40}. The upper and lower panels
  correspond to the relativistic and heavy-baryon approach,
  respectively. The light and dark-shaded bands (the latter nearly shrink
  to lines)  refer to
  the deltaless and deltafull calculations, respectively. The bands correspond to using the KH and GW sets of LECs $c_i$.}
\label{fig:sigmadiffdOdT}
\end{figure}

The heavy-baryon approach also fails to describe various
differential cross sections at NLO, most noticeable the
double-differential cross section with respect to the solid angle
$\Omega_2$ and the pion kinetic energy $T_2=\omega_2 -M_\pi$ 
in the channel $\pi^-p\to\pi^+\pi^-n$. The data
for this observable are reported in Ref.~\cite{Manley:1984zs} and the
comparison between the predictions of relativistic and heavy-baryon
$\chi$PT are presented in Fig.~\ref{fig:sigmadiffdOdT}. 
While the
inclusion of the delta in the HB formulation at order $\epsilon^2$
shifts the theoretical results towards the data, these shifts are too
small and unable to bring the theory in agreement with the data.  
We emphasize,
however, that the data are well described by the 
next-to-next-to-leading order ($Q^3$) deltaless HB 
calculation  of
Ref.~\cite{Fettes:1999wp}. The
relativistic deltafull approach is able to describe the data 
properly already at NLO. Moreover, even the deltaless covariant
formulation yields a reasonably good description of the data at
this order. The description of the data is somewhat better in
$\chi$PT with explicit delta except for the cross section at
$\sqrt{s}=1262$ MeV and the largest value of the kinetic energy of
$\pi^+$, $T_2=31.4$ MeV. 

\begin{figure}[p]
  \centering
\subfigure{
 \includegraphics[width=0.95\textwidth]{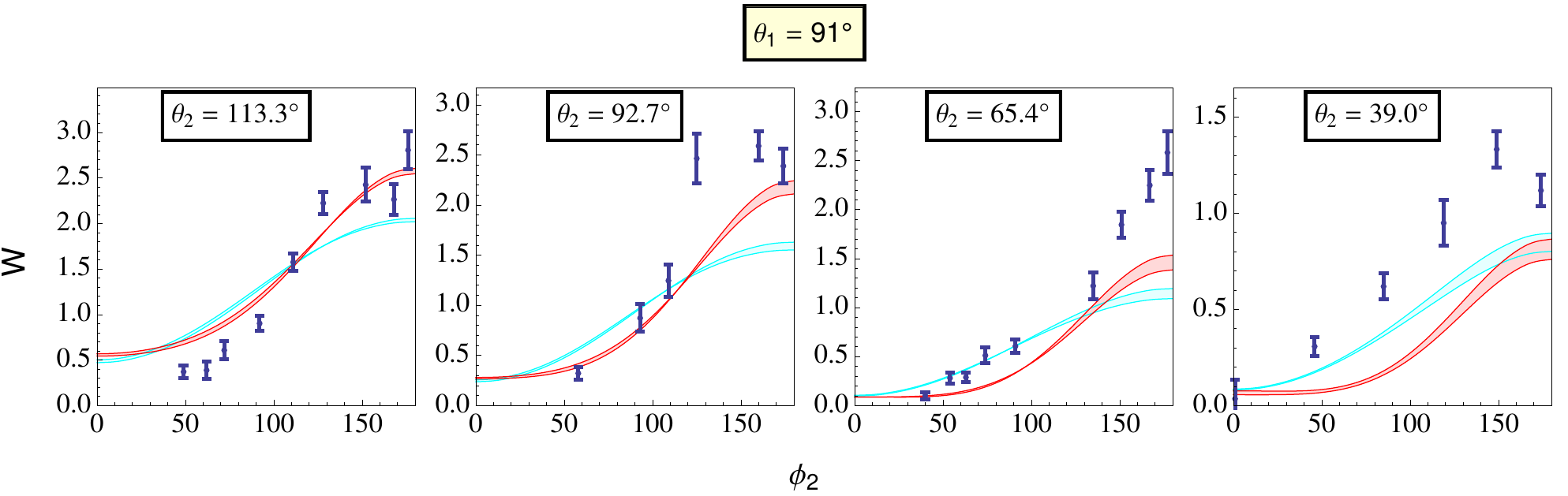}}
\vskip -13pt
\subfigure{
 \includegraphics[width=0.95\textwidth]{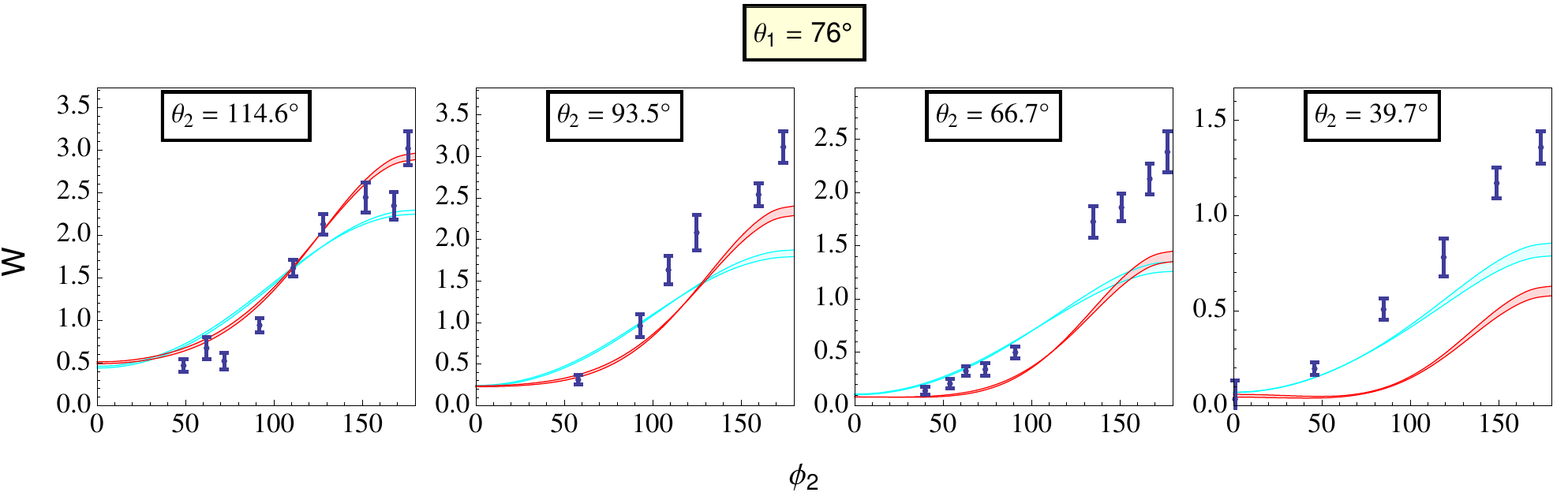}}
\vskip -13pt
\subfigure{
 \includegraphics[width=0.95\textwidth]{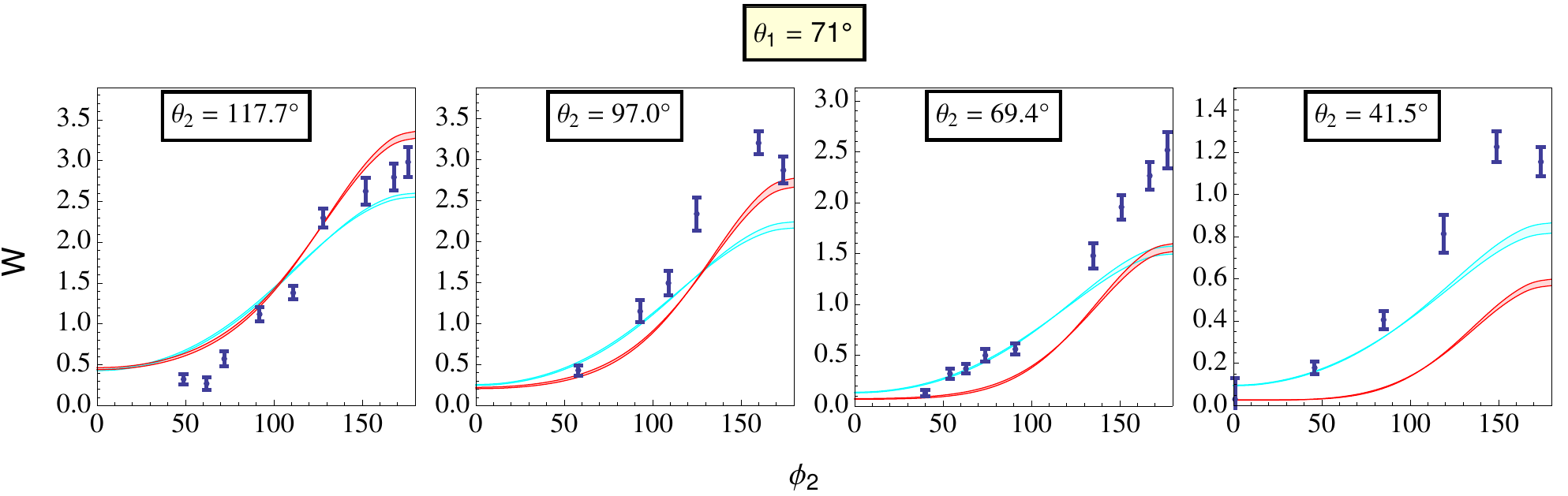}}
\vskip -13pt
\subfigure{
 \includegraphics[width=0.95\textwidth]{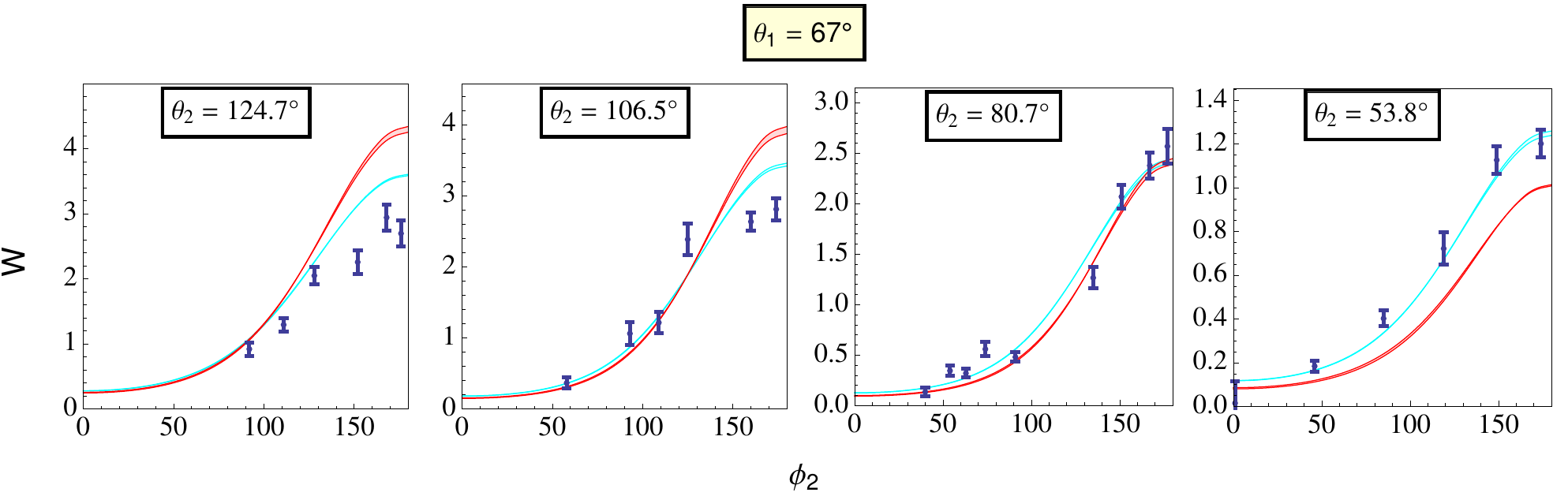}}
  \caption{NLO relativistic $\chi$PT predictions for the angular correlation functions
  in the $\pi^-p\to\pi^+\pi^-n$ channel at fixed $\theta_1$ and
  $\theta_2$ for
$\sqrt{s}=\unit{1301}{\mega\electronvolt}$, see Eq. \eqref{eq:45}. 
 The light and dark-shaded bands refer to
  the deltaless and deltafull calculations, respectively. The bands
  correspond to using the KH and GW sets of LECs $c_i$.}
  \label{fig:sigmadiffW1}
\end{figure}

\begin{figure}[p]
  \centering
\subfigure{
 \includegraphics[width=0.74\textwidth]{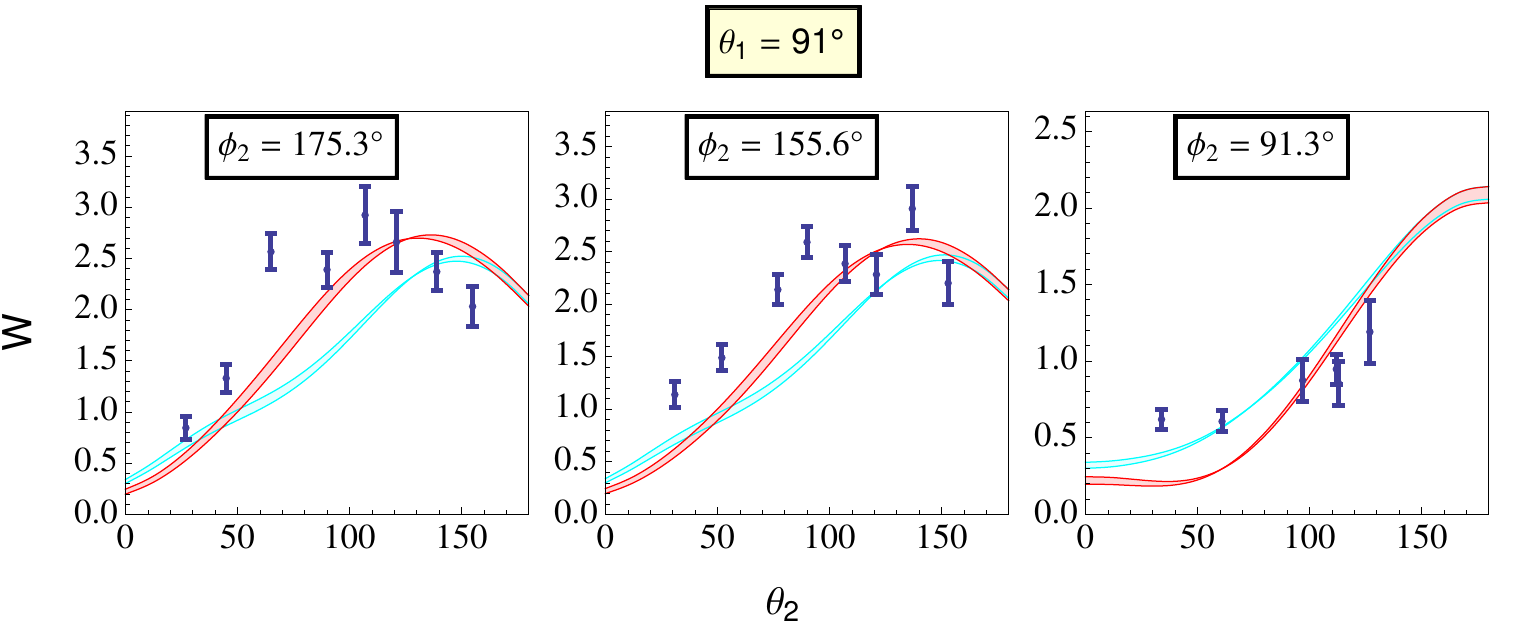}}
\vskip -13pt
\subfigure{
 \includegraphics[width=0.74\textwidth]{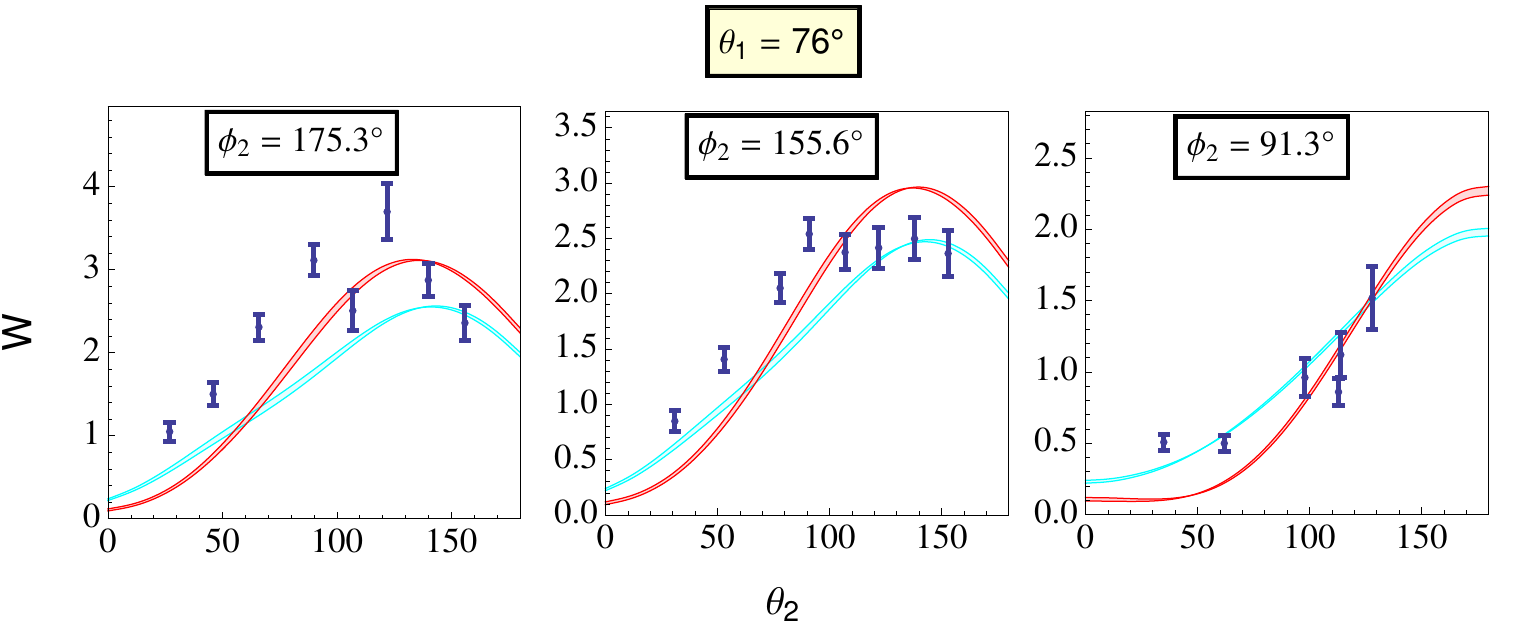}}
\vskip -13pt
\subfigure{
 \includegraphics[width=0.74\textwidth]{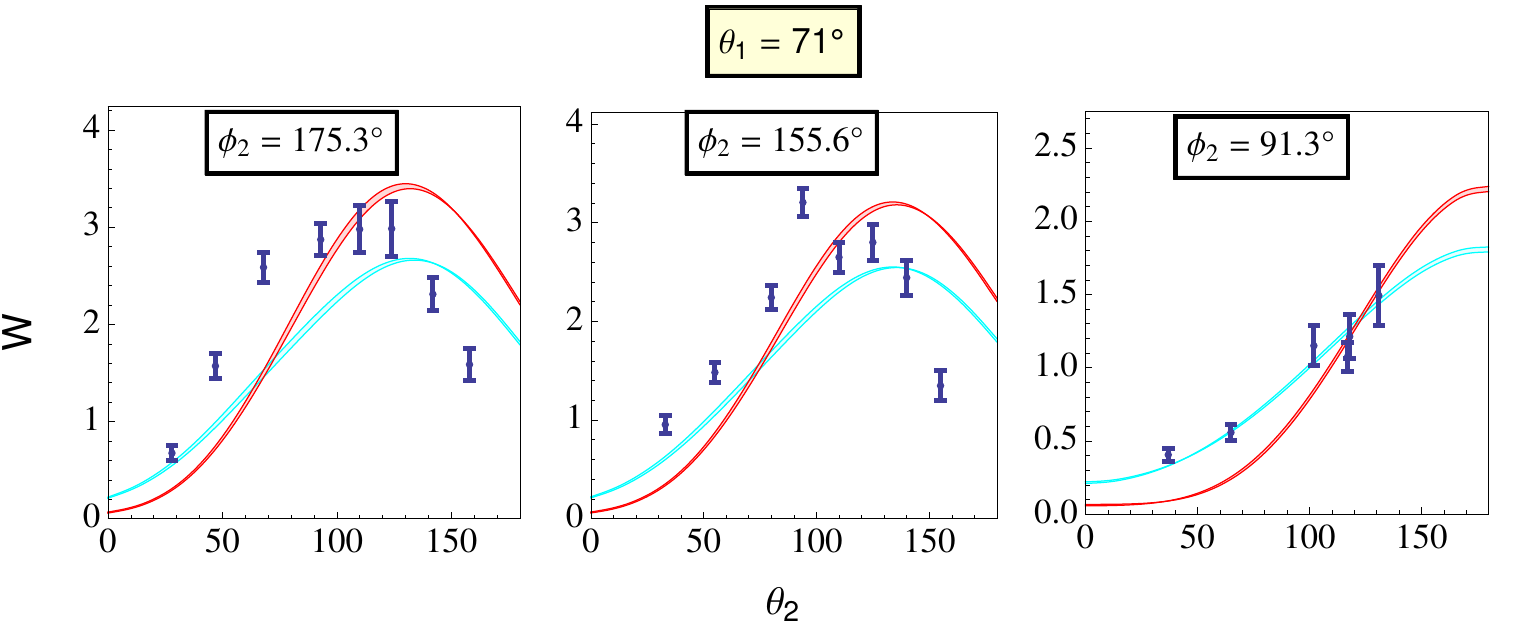}}
\vskip -13pt
\subfigure{
 \includegraphics[width=0.74\textwidth]{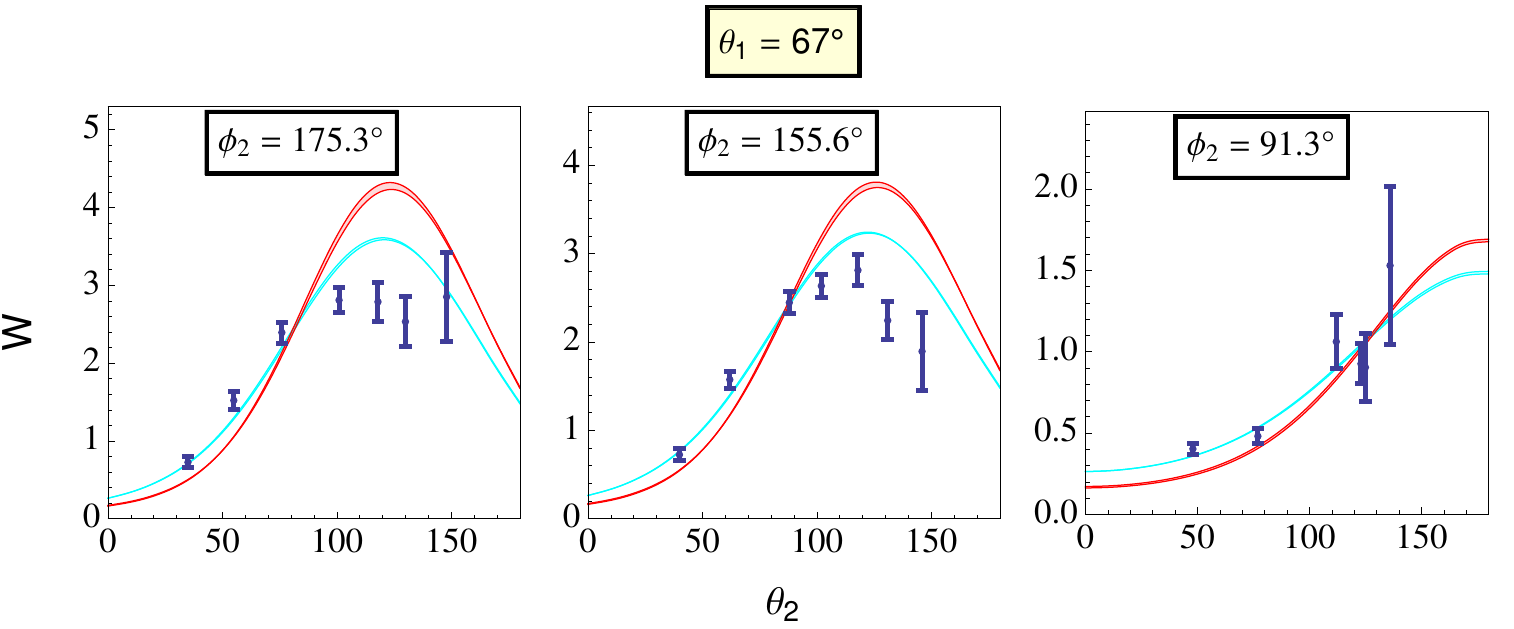}}
\vskip - 0.3 true cm 
  \caption{Comparison of NLO relativistic deltafull and deltaless $\chi$PT  predictions for the angular correlation functions
  in the $\pi^-p\to\pi^+\pi^-n$ channel at fixed $\theta_1$ and
  $\phi_2$ for
$\sqrt{s}=\unit{1301}{\mega\electronvolt}$, see Eq. \eqref{eq:45}. For
remaining notation see Fig.~\ref{fig:sigmadiffW1}. }
  \label{fig:sigmadiffW2}
\end{figure}

Given the failure of the NLO HB approach for the double-differential
cross sections, we will leave out the HB
predictions  in what follows and focus entirely on the results based on
the relativistic framework. First, we consider the angular
correlation function $W$ defined in
Eq.~(\ref{eq:45}) in the channel $\pi^- p \to \pi^+ \pi^- n$. 
Fig.~\ref{fig:sigmadiffW1}  (Fig.~\ref{fig:sigmadiffW2})
show our NLO predictions in the deltaless and deltafull relativistic
approaches in comparison with experimental data taken from 
Ref.~\cite{Muller:1993pb}  for fixed $\theta_1$ and $\theta_2$
($\theta_1$ and $\phi_2$).  One observes that the deltafull results
tend to have a stronger curvature which, in most cases, is in a better qualitative
agreement with the shape of the experimental data. Generally, the data are reasonably
well described in both approaches (given that the calculations are
carried out at NLO in the low-energy expansion). The largest
deviations between the $\epsilon^2$ results and  the data emerge 
at lowest values of $\theta_2$ and $\theta_1=39.0\ldots 41.5^\circ$, 
see Fig.~\ref{fig:sigmadiffW1}. In fact, the predictions of the 
deltaless framework appear to be closer to the data in these cases
(although the shape of the data is better described in the deltafull
theory). Given that these cases correspond to the largest differences 
between the $Q^2$ and $\epsilon^2$ results, and their magnitude is
comparable with the deviation from the data, it is conceivable that 
higher-order corrections  
might be significant in these kinematical conditions.   

\begin{figure}[p]
  \centering
\subfigure{
 \includegraphics[width=0.485\textwidth]{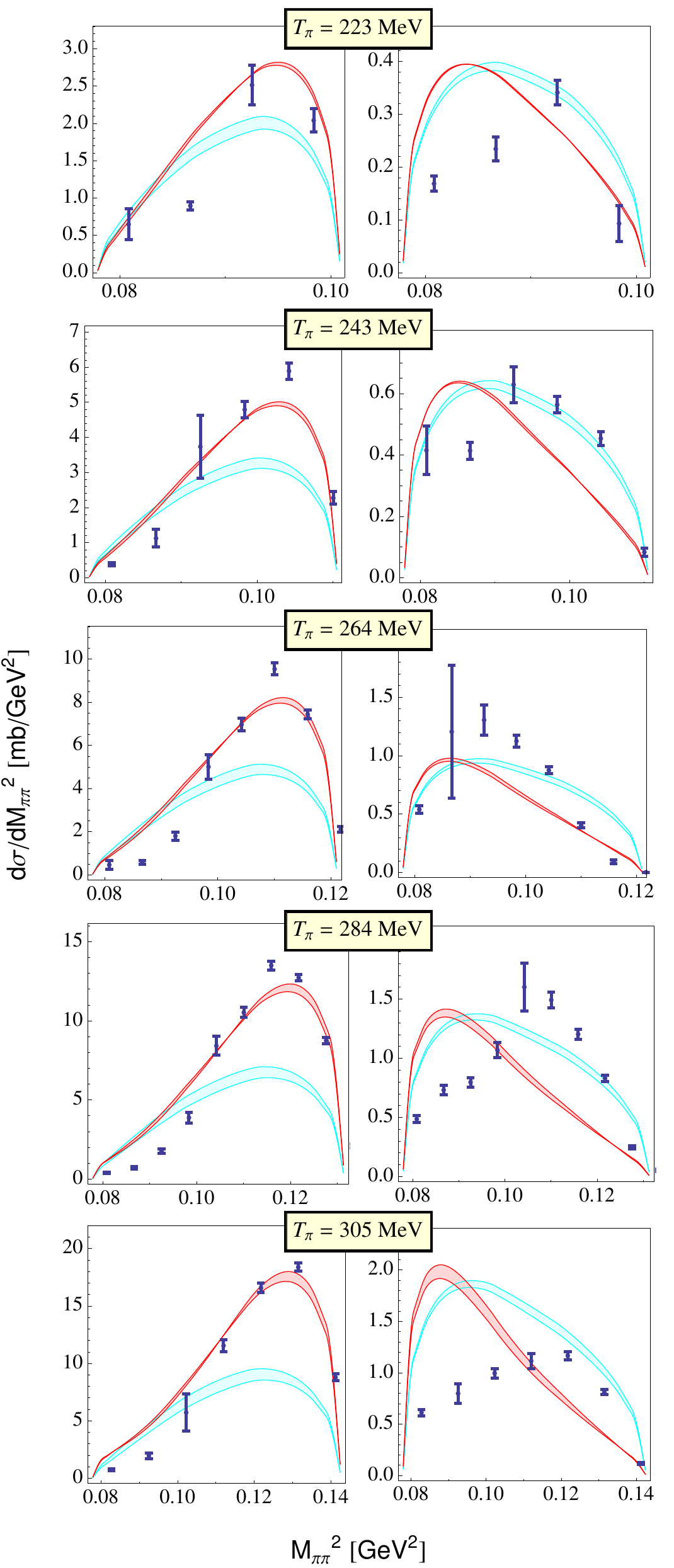}}
\subfigure{
 \includegraphics[width=0.485\textwidth]{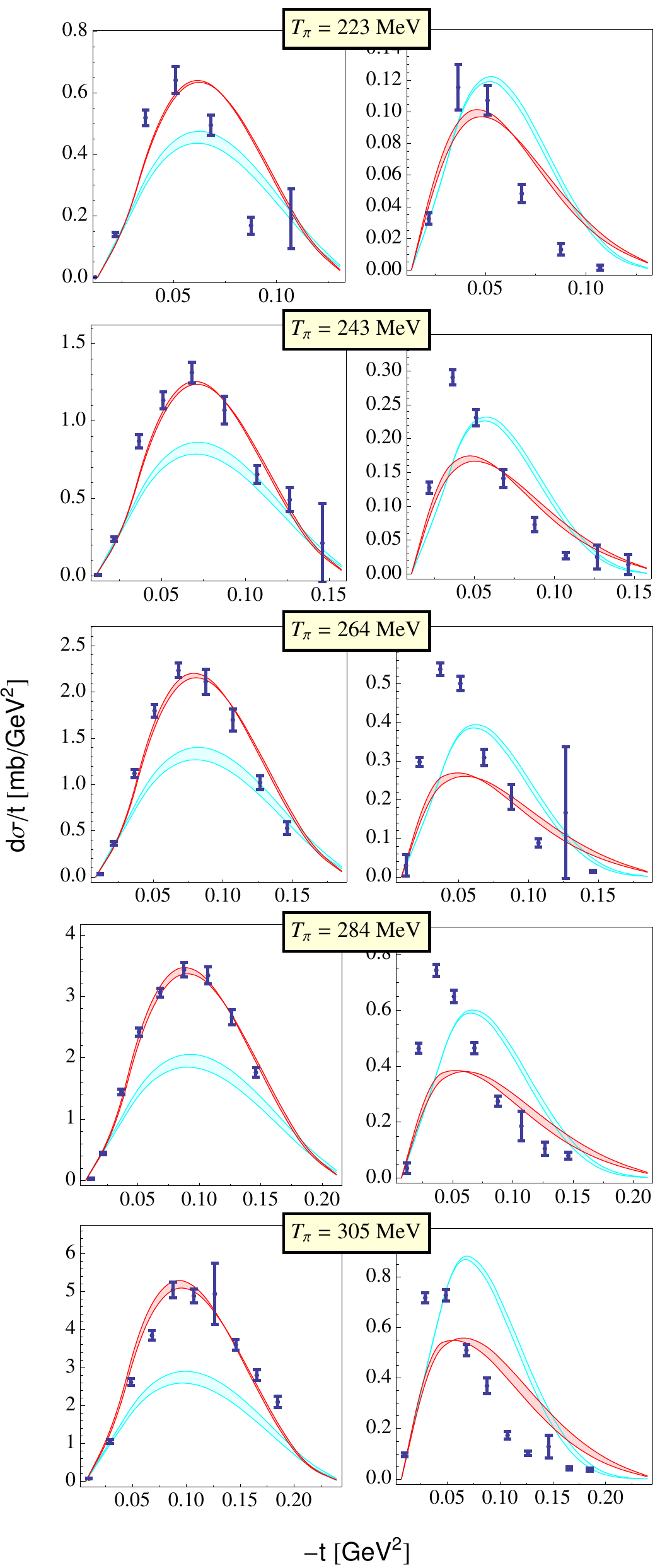}}
  \caption{Comparison of NLO relativistic deltafull and deltaless $\chi$PT  predictions 
    for the single-differential cross sections with respect to $M_{\pi\pi}^2$
    and $t$, respectively, between the two channels
    $\pi^-p\to\pi^+\pi^-n$ (left) and  $\pi^+p\to\pi^+\pi^+n$ (right), see Eq. \eqref{eq:47}.  For
remaining notation see Fig.~\ref{fig:sigmadiffW1}.}
  \label{fig:sigmadiffdManddt}
\end{figure}

\begin{figure}[p]
  \centering
\subfigure{
 \includegraphics[width=0.48\textwidth]{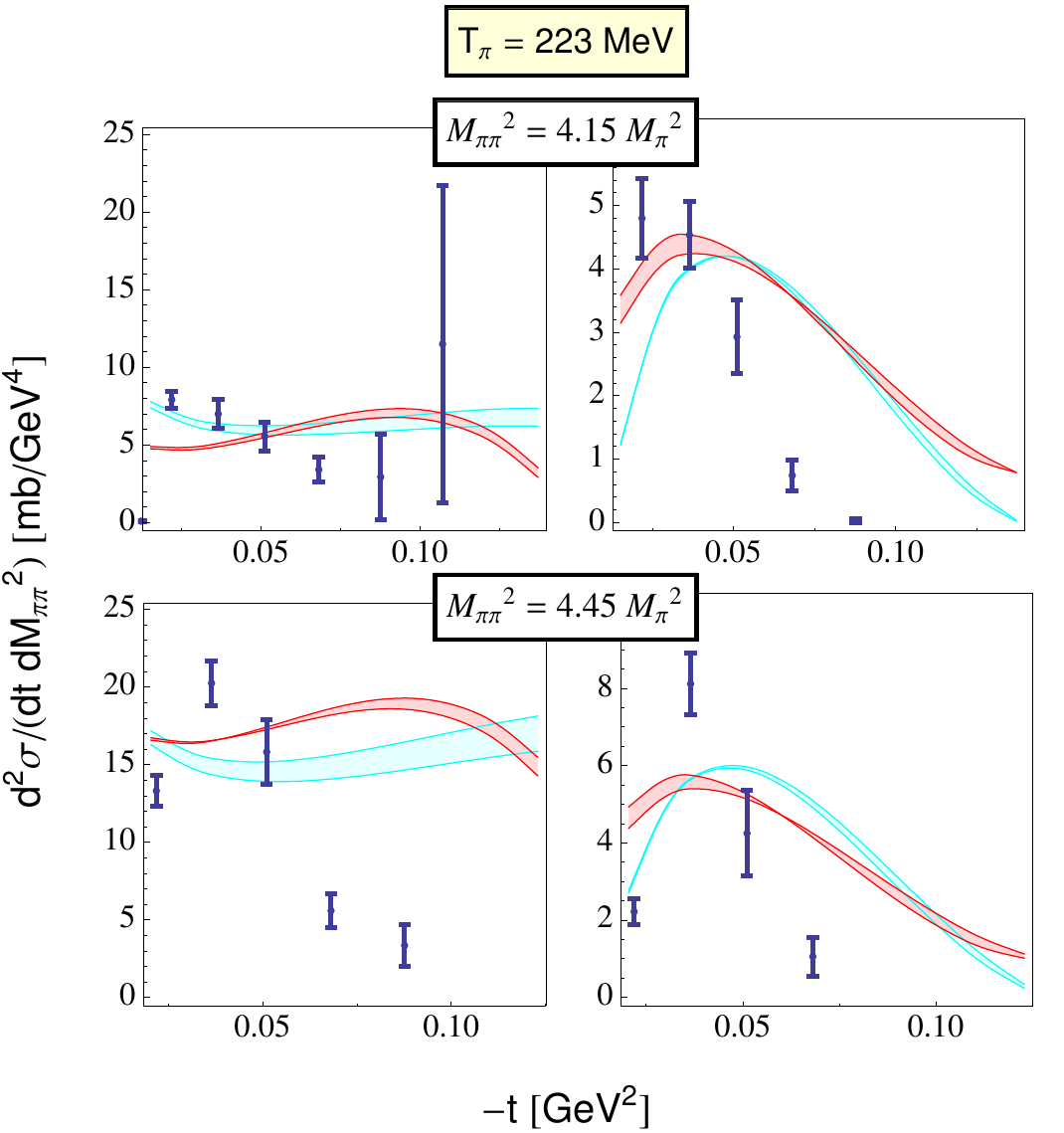}}
\subfigure{
 \includegraphics[width=0.48\textwidth]{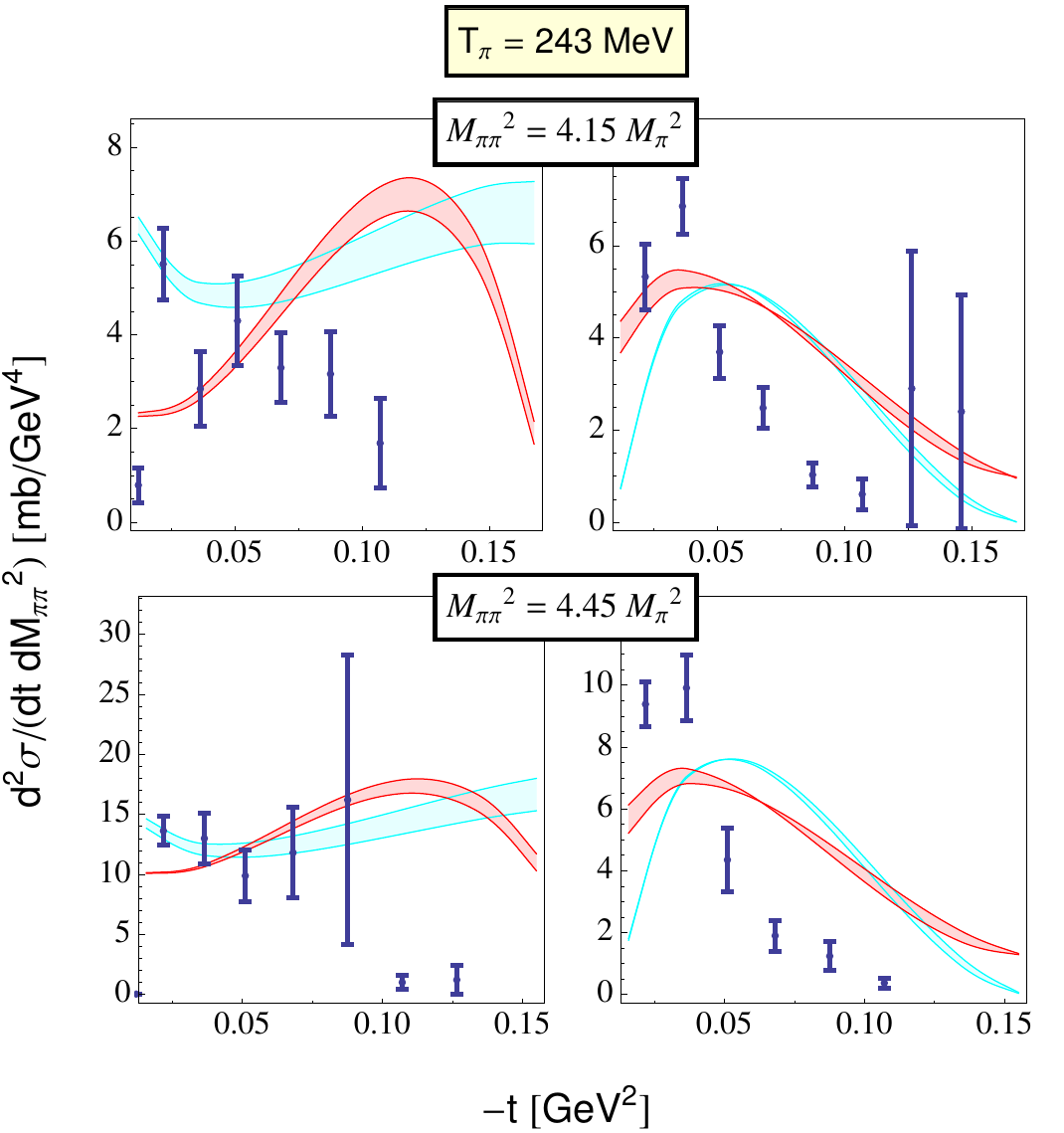}}
  \caption{Comparison of NLO relativistic deltafull and deltaless $\chi$PT  predictions 
    for the double-differential cross sections with respect to $M_{\pi\pi}^2$
    and $t$ between the two channels
    $\pi^-p\to\pi^+\pi^-n$ (left) and  $\pi^+p\to\pi^+\pi^+n$ (right)
    for two different incoming pion energies, see Eq. \eqref{eq:47}.  For
remaining notation see Fig.~\ref{fig:sigmadiffW1}.}
  \label{fig:sigmadiffdMdt}
\end{figure}

We next turn to the single-differential cross sections with respect to $M_{\pi\pi}^2$ and
 $t$, see Eq.~(\ref{eq:47}). Our predictions for
 $d\sigma/dM_{\pi\pi}^2$ and $d\sigma / dt$ are shown in comparison
 with the experimental data from Ref.~\cite{Kermani:1998gp} in the
 left and right panels of Fig.~\ref{fig:sigmadiffdManddt}, respectively.  
In each case we compare the two channels, namely
$\pi^-p\to\pi^+\pi^-n$ and $\pi^+p\to\pi^+\pi^+n$.  
We recall that the total cross section is very accurately predicted 
at order $\epsilon^2$ in the $\pi^+ \pi^-$ channels, see Fig.~\ref{fig:sigmatot}. It is comforting to see that 
both single-differential cross sections are also well described in the
deltafull approach. On the other hand, deltaless results at 
order $Q^2$ strongly underpredict the experimental data which is in
line with the observed underprediction of the total cross section.  
This is the most pronounced example of the importance of the 
explicit inclusion of the delta isobar we found in our analysis. 
In the $\pi^+\pi^+$-channel, 
the single-differential cross sections
are found to be poorly described at both $Q^2$ and $\epsilon^2$ orders. In
particular, even the shape of the cross section $d\sigma/dM_{\pi\pi}^2$ 
is not correctly reproduced. The situation is slightly better for 
the cross section $d \sigma /d t$. The observed large deviations from
the data should not come as a surprise given that the predicted total
cross section  significantly overestimates the experimental data both
in the deltaless and deltafull formulations. 
We also looked at the double differential
cross section $d^2\sigma /(d t d M_{\pi \pi})$ in the same two
channels and found large deviations between the theory and the data,
see Fig.~\ref{fig:sigmadiffdMdt}.   The large discrepancies between
the theory and experimental data in the $\pi^+\pi^-$-channel,
where  the single-differential and total cross sections are well
reproduced, appear to be somewhat surprising.  

\begin{figure}[p]
  \centering
\subfigure{
 \includegraphics[width=0.53\textwidth]{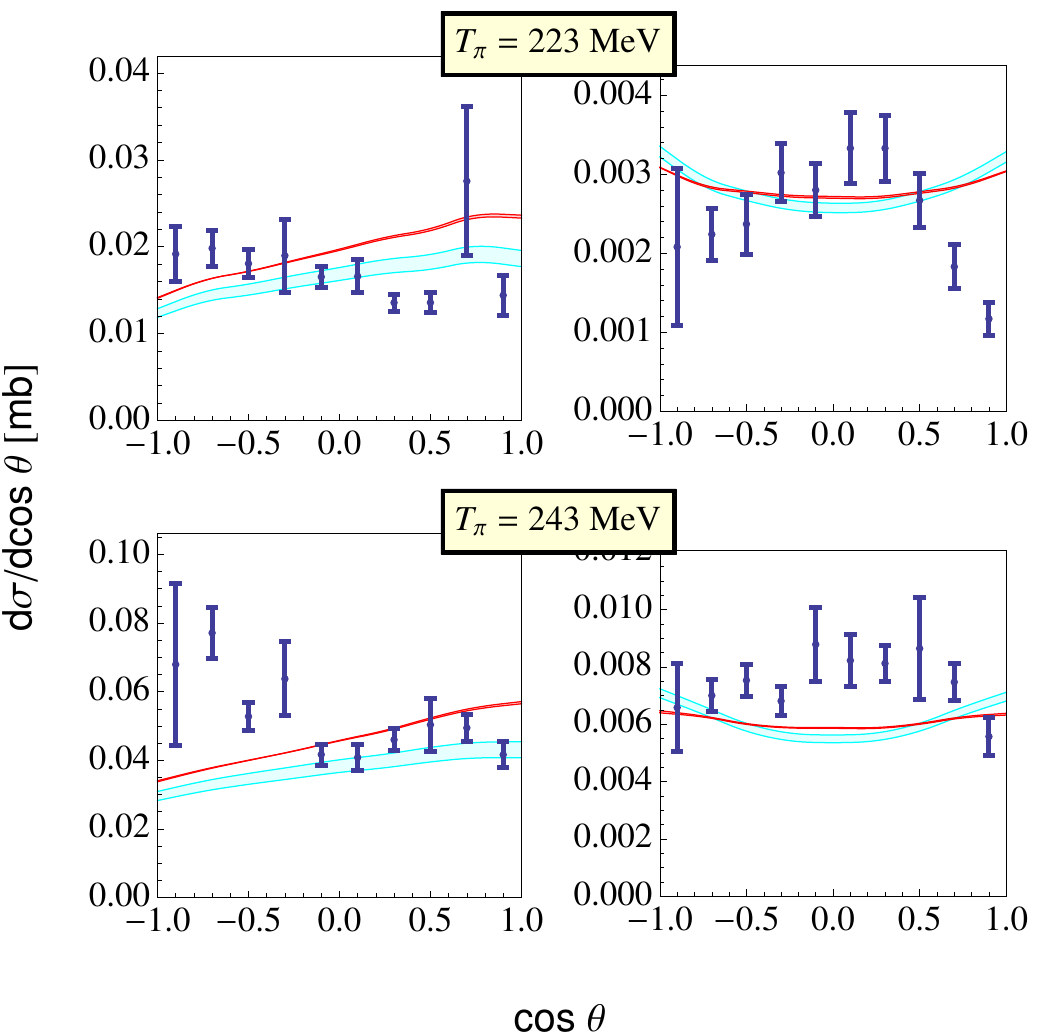}}
  \caption{Comparison of NLO relativistic deltafull and deltaless $\chi$PT   predictions 
    for the single-differential cross sections with respect to $\cos\theta$ between the two channels
    $\pi^-p\to\pi^+\pi^-n$ (left) and  $\pi^+p\to\pi^+\pi^+n$ (right)
    for two different incoming pion energies, see Eq. \eqref{eq:47}. For
remaining notation see Fig.~\ref{fig:sigmadiffW1}.}
  \label{fig:sigmadiffdcosth}
\end{figure}

Finally, our results for the single-differential cross sections with
respect to $\cos \theta$ at two lowest energies in the
$\pi^-p\to\pi^+\pi^-n$ and $\pi^+p\to\pi^+\pi^+n$
channels are visualized in Fig.~\ref{fig:sigmadiffdcosth}. Our
predictions have the same magnitude as the experimental data, but show
a different shape. As might be expected from the results for the total
cross section, the deviations are most pronounced in the
$\pi^+\pi^+$-channel. Notice further that the effect of the explicit
treatment of the delta isobar is fairly minor for these particular observables.  

It is instructive to compare our results with the earlier calculations
within the HB \cite{Fettes:1999wp}  and relativistic \cite{Bernard:1997tq}
frameworks.   For the total cross sections, our HB results at LO and
NLO are very close to the corresponding ones of
Ref.~\cite{Fettes:1999wp} and feature similar underprediction of the
data for the $\pi^- p$ case. The $Q^3$ results of that work show a
significant improvement, which we now interpret as resulting mainly
from taking into account $1/m_N$ corrections. Also for the double
differential cross sections $d^2\sigma / d\Omega_2 dT_2$, our
predictions at  $\sqrt{s} = 1242$ MeV at LO and NLO and  $\sqrt{s} =
1242$ MeV at LO  are  close to the ones of Ref.~\cite{Fettes:1999wp}. 
Our NLO results at the higher energy appear, however, to be significantly  closer to the 
data than the ones of that work which probably can be traced back to
the different choice of $c_i$'s.  While not explicitly shown, we
observe that our HB results for other observables are similar to the
ones of \cite{Fettes:1999wp}. Finally, the order-$Q^2$ relativistic
$\chi$PT calculation of 
Ref.~\cite{Bernard:1997tq} also provides a very useful benchmark for our analysis. 
We have verified that our results agree with the ones for all
observables shown in that work.  Finally, we compare with the results of Ref.~\cite{Jensen:1997em}.
Their leading order calculation with explicit deltas and the Roper describes the total
cross section data somewhat better than our LO deltafull approach, however, it is of
similar quality to our NLO deltafull calculation. 
It is conceivable that this difference is mostly due to the explicit inclusion
of the Roper resonance in Ref.~\cite{Jensen:1997em}. Note, however, that there is no power counting
underlying this calculation as it only uses the leading dimension one
 derivative pion-baryon couplings.

\section{Summary and outlook}
\label{sec:sum}

In this paper we have analyzed single pion production off nucleons at tree
level up to NLO using the heavy-baryon and manifestly covariant
formulations of $\chi$PT with and without inclusion of explicit delta
isobar degrees of freedom. The main results of our study can be
summarized as follows:
\begin{itemize}
\item
We worked out the leading and subleading contributions of the delta
isobar to the invariant amplitudes in the reaction $\pi N\to\pi\pi N$
using both the HB and manifestly covariant formulations of $\chi$PT.  
\item
In order to determine the low-energy constants $b_i$ entering the
subleading pion-nucleon-delta Lagrangian, several global fits to the
available low-energy data for the total cross sections in the five
channels $\pi^- p \to \pi^0 \pi^0 n$,  $\pi^- p \to \pi^+ \pi^- n$,  
$\pi^+ p \to \pi^+ \pi^+ n$,  $\pi^+ p \to \pi^+ \pi^0 p$ and 
$\pi^- p \to \pi^0 \pi^- p$ have been performed. For the LECs $c_i$,
which parametrize subleading pion-nucleon interactions, we adopted the values
extracted from pion-nucleon scattering.  Using the large-$N_c$
predictions for the LECs $h_A$ and $g_1$ entering the LO Lagrangians
$\mathcal{L}_{\pi N \Delta}^{(1)}$  and $\mathcal{L}_{\pi N
  \Delta}^{(1)}$ and restricting the energy in the fit by 
$T_\pi=250$ MeV, the extracted values for the linear combinations $b_4+b_5$  and
$b_3 + b_6$ are found to be of a natural size when using the
covariant approach. We observe strong anticorrelations between the
LECs $b_4$ and $b_5$ as well as $b_3$ and $b_6$ which prevent a
reliable determination of the linear combinations  $b_4-b_5$  and $b_3
- b_6$. The anticorrelations are found to be much less pronounced if the energy range in the
fit is increased up to $T_\pi = 400$ MeV. The resulting values of
all LECs $b_i$ are then found to be of a reasonably natural size. This is
in contrast with the fits carried out in the HB approach, where an
unnaturally large value for $b_4 + b_5$ is found. Notice that in this
formulation the amplitude does not depend on the LECs $b_{3,6}$, and the LECs
$b_{4,5}$ are also found to be strongly anticorrelated. 
\item
We explored the sensitivity of the total cross section to the
LO $\pi \Delta \Delta$ coupling $g_1$, which is difficult to access in
other processes such as e.g.~pion-nucleon scattering, by performing
unconstrained fits to the total cross section data. The resulting values $g_1 =
1.36 \pm 0.73$ and $g_1=1.68 \pm 1.38$ in the HB and relativistic
formulations, respectively, are (in the HB approach only barely) consistent  with
the the large-$N_c$ prediction for this LEC, namely $g_1 =2.27$. Extending the fit to
$400$ MeV within the relativistic formulation leads to a somewhat smaller
value of $g_1=1.41 \pm 0.22$.     
\item
We found that the covariant framework allows for a clearly superior
description of the experimental data at NLO as compared to
the HB formulation at the same order. Further, as expected, the
explicit treatment of the delta isobar leads to a better 
description of the data compared to the standard deltaless
formulation, most notably of the $\pi^- p \to \pi^0 \pi^0 n$ and $\pi^-
p \to \pi^+ \pi^- n$ total cross sections at  higher energy as well as
of the single-differential cross sections with respect to $M_{\pi\pi}$
and $t$ in the $\pi^- p \to \pi^+ \pi^- n$ channel. Still, certain
single- and double-differential cross sections could not be properly
described at this order in the chiral expansion. Finally, we found
that there is fairly minor dependence of the extracted LECs and
predictions for various observables on the variation in the LECs $c_i$
used as input in our calculation.  
\end{itemize}
In the future, this work has to be extended in several
directions. First of all, one has to go to next-higher order in the
chiral expansion and include pion loop contributions within
the covariant framework. This will not only allow one to test the
convergence of the chiral expansion, but also possibly constrain 
the LEC $d_{16}$ which governs the quark mass dependence of the
nucleon axial vector coupling constant. In addition, it would be very
interesting to carry out a simultaneous analysis of the reactions $\pi
N \to \pi N$ and $\pi N \to \pi \pi N$. We expect that such a study
will result in  a more precise determination of the LECs $c_i$ as
compared to elastic pion-nucleon scattering. These LECs govern, in
particular, the longest-range three-nucleon force and thus play a prominent
role in ongoing studies of few- and many-nucleon systems. It is also
conceivable that such a combined analysis will allow for a better
determination of the LO pion-delta coupling constant $g_1$. As a further
step, the explicit inclusion of the Roper resonance might also be 
considered. Work along these lines is in progress.

\section*{Acknowledgments}

We would like to thank Igor Strakovsky for useful comments on the
manuscript. This work was supported by the DFG (SFB/TR 16,
``Subnuclear Structure of Matter''), the European
Community-Research Infrastructure Integrating Activity ``Study of
Strongly Interacting Matter'' (acronym HadronPhysics3,
Grant Agreement n. 283286) under the Seventh Framework Programme of EU,
and the ERC project 259218 NUCLEAREFT.

\clearpage
\appendix
\section{Kinematics and weight functions}
\label{sec:KinWeiFunc}
The weight functions $y_{ij}$ appearing in Eq. \eqref{eq:31} are
defined as follows
\begin{equation}
  \label{eq:192}
  \begin{aligned}
    y_{11}&= -m_N^2+M_\pi^2-s+s_1+s_2-t_1-t_2 \komma\\
    y_{12}&= -m_N \left(2 m_N^2+2 s-2 s_1-2 s_2+t_1+t_2\right) \komma\\
    y_{13}&= m_N (-t_1+t_2) \komma\\
    y_{14}&= s (s_1-s_2)+s_2 (s_2-t_1)+M_\pi^2 (t_1-t_2)+m_N^2
    (s_1-s_2+t_1-t_2)+s_1 (-s_1+t_2) \komma\\
    y_{22}&= -3 m_N^4+2 M_\pi^4-M_\pi^2 (3 s+s_1+s_2)+s
    (s+t_1+t_2) \\
    &\qquad-m_N^2 \left(3 M_\pi^2+6 s-4 s_1-4 s_2+t_1+t_2\right) \komma\\
    y_{23}&= s s_1-s s_2+s_1 t_1-s_2 t_2+m_N^2
    (-s_1+s_2-t_1+t_2)+M_\pi^2 (-2 s_1+2 s_2-t_1+t_2) \komma\\
    y_{24}&= m_N \left(m_N^2+2 M_\pi^2+s-s_1-s_2\right) (2 s_1-2
    s_2+t_1-t_2) \komma\\
    y_{33}&= 3 m_N^4-2 M_\pi^4-(s-2 s_1) (s-2 s_2+t_1)-(s-2 s_2) t_2 \\
    &\qquad-m_N^2 \left(5 M_\pi^2-2 s+2 s_1+2
      s_2+t_1+t_2\right)+M_\pi^2 (3 s-3 s_1-3 s_2+2 (t_1+t_2)) \komma\\
    y_{34}&= -m_N \left(m_N^2-2 M_\pi^2+s-s_1-s_2\right) \left(2
      m_N^2+4 M_\pi^2-2 s-t_1-t_2\right) \komma\\
    y_{44}&= \left(m_N^2+s-s_1-s_2\right) \left((s_1-s_2)
      (-s_1+s_2-t_1+t_2)+\left(-m_N^2+s\right)
      \left(-m_N^2+s+t_1+t_2\right)\right) \\
    &\qquad+M_\pi^2 \left(-5 s^2+4 M_\pi^2 \left(-M_\pi^2+2
        s\right)-s_1^2+6 s_1 s_2-s_2^2+2 (s_1-s_2) (-t_1+t_2)\right.\\
      &\qquad\left.+m_N^2 \left(3 m_N^2-8 M_\pi^2+6 (s-s_1-s_2)+2 (t_1+t_2)\right)+2 s(s_1+s_2-t_1-t_2)\right),
  \end{aligned}
\end{equation}
where $y_{ij}=y_{ji}$ and the products of the four vectors are expressed  in
terms of the Mandelstam variables via
\begin{equation}
  \label{eq:202}
  \begin{aligned}
    2 p\cdot q_1 &= s-m_N^2-M_\pi^2  \komma
   & 2p\cdot q_2 &= s-s_2+t_1-M_\pi^2 \komma\\
    2p\cdot q_3 &= s-s_1 +t_2 -M_\pi^2 \komma
   & 2p\cdot p^\prime &= s_1 + s_2 -s -t_1 -t_2 +m_N^2 +M_\pi^2\komma\\ 
   2 q_1\cdot q_2 &= 2M_\pi^2 -t_1 \komma
   &2 q_1\cdot q_3 &= 2M_\pi^2 -t_2\komma \\
   2 q_1\cdot p^\prime &= s+t_1+t_2-m_N^2-3M_\pi^2\komma
   &2 q_2 \cdot q_3 &= s -s_1 -s_2 +m_N^2\komma\\
   2q_2\cdot p^\prime &= s_1 -m_N^2 -M_\pi^2\komma
   &2q_3\cdot p^\prime &= s_2 -m_N^2 -M_\pi^2\punkt
  \end{aligned}
\end{equation}


\pagebreak


$\,$

\section*{Figures}

\vspace{1.5cm}

\vskip 2 true cm

\bigskip

\vskip 2 true cm

\bigskip 

\end{document}